\documentclass[USenglish,onecolumn,openany]{article}

\usepackage[utf8]{inputenc}%(only for the pdftex engine)
\usepackage[big]{dgruyter}

\usepackage{amsmath}
\usepackage{graphicx}
\usepackage{enumerate}

\usepackage{mathtools,amssymb}
\usepackage{multicol}
\usepackage{indentfirst}
\usepackage{float}
\usepackage{algorithm}
\usepackage{algorithmicx}
\usepackage{algpseudocode} % REQUIRED for \State, \For, etc.

\usepackage{longtable}
\usepackage{setspace}
\usepackage{amsthm}
\usepackage{tcolorbox}
\usepackage{caption}
\usepackage{tabularx}
\begin{document}
 \sloppy

  \articletype{Research Article{\hfill}Open Access}

  \author*[1]{Lauren Eyler Dang}

\author[2]{Jens Magelund Tarp}

\author[3]{Trine Julie Abrahamsen}

\author[4]{Kajsa Kvist}

\author[5]{John B Buse}

\author[6]{Maya Petersen}

\author[7]{Mark van der Laan}

   \affil[1]{Biostatistics Research Branch, National Institute of Allergy and Infectious Diseases, Rockville, MD, USA, 20852; E-mail: lauren.dang@nih.gov}

  \affil[2]{Novo Nordisk, S$\o$borg, Denmark; E-mail: jqmt@novonordisk.com}

  \affil[3]{Novo Nordisk, S$\o$borg, Denmark; E-mail: tjla@novonordisk.com}

  \affil[4]{Novo Nordisk, S$\o$borg, Denmark; E-mail: tekk@novonordisk.com}

  \affil[5]{Division of Endocrinology, Department of Medicine, University of North Carolina, Chapel Hill, NC, USA, 27516; E-mail: john\_buse@med.unc.edu}

\affil[6]{Department of Biostatistics, University of California, Berkeley, CA, USA, 94720 ; E-mail: mayaliv@berkeley.edu}

\affil[7]{Department of Biostatistics, University of California, Berkeley, CA, USA, 94720 ; E-mail: laan@berkeley.edu}

  \title{\huge Experiment-selector cross-validated targeted maximum likelihood estimator for hybrid RCT-external data studies}

  \runningtitle{Experiment-Selector CV-TMLE}
    \runningauthor{Lauren Eyler Dang et al.}

  %\subtitle{...}

  \begin{abstract}
{Augmenting a randomized controlled trial (RCT) with external data may increase power at the risk of introducing bias. To select and analyze the experiment (RCT alone or combined with external data) with the optimal bias-variance tradeoff, we develop a novel experiment-selector cross-validated targeted maximum likelihood estimator for randomized-external data studies (ES-CVTMLE). This estimator utilizes two estimates of bias to determine whether to integrate external data based on 1) a function of the difference in conditional mean outcome under control between the RCT and combined experiments and 2) an estimate of the average treatment effect on a negative control outcome (NCO). We define the asymptotic distribution of the ES-CVTMLE under varying magnitudes of bias and construct confidence intervals by Monte Carlo simulation. We evaluate ES-CVTMLE compared to three other data fusion estimators in simulations and demonstrate the ability of ES-CVTMLE to distinguish biased from unbiased external controls in a real data analysis of the effect of liraglutide on glycemic control from the LEADER trial. The ES-CVTMLE has the potential to improve power while providing relatively robust inference for future hybrid RCT-external data studies.}
\end{abstract}
  \keywords{causal inference, clinical trials, data fusion, negative control outcomes, real world data}
   \classification[MSC]{62G05}
 % \communicated{...}
 % \dedication{...}

\DOI{DOI}
  \startpage{1}
  \received{..}
  \revised{..}
  \accepted{..}

  \journalyear{2024}
  \journalvolume{1}
%  \journalissue{1}

\maketitle

\section{Introduction} \label{intro}
With the growing availability of observational data from sources such as registries, electronic health records, or the control arms of previous trials, the power of randomized controlled trials (RCTs) could potentially be improved while randomizing fewer participants to control status if we were able to incorporate data from these sources in the analysis \citep{pocockCombinationRandomizedHistorical1976,vieleUseHistoricalControl2014, schmidliRobustMetaanalyticpredictivePriors2014}. Yet combining RCT with external data comes with the risk of introducing bias from multiple sources, including measurement error, selection bias, and confounding \citep{bareinboimCausalInferenceDatafusion2016}. If we were to simply pool the RCT and external data, we could potentially have large power gains, but we would have to rely on strong assumptions to conclude that a causal effect had been estimated. Analysis of the RCT alone would allow for estimation of a causal effect by design, but in some contexts, running an adequately-powered trial is infeasible \citep{jahanshahiUseExternalControls2021}. \par
In such cases, it may be valuable to consider an approach in between the extremes of a) pooling RCT and external data and b) only utilizing RCT data. Data fusion estimators aim to fill this gap; instead of relying on untestable causal assumptions, they aim to estimate the bias that may be introduced by incorporating external data in order to decide whether to include external data or how to weight external data in a hybrid RCT-observational analysis. For example, Bayesian dynamic borrowing approaches generate a prior distribution of the RCT control parameter based on external control data. These methods take different approaches to down-weighting the observational information based on heterogeneity between RCT and external control outcomes \citep{pocockCombinationRandomizedHistorical1976, ibrahimPowerPriorDistributions2000, hobbsCommensuratePriorsIncorporating2012, schmidliRobustMetaanalyticpredictivePriors2014}. Taking a Frequentist approach, multiple estimators conduct a test to decide whether to incorporate external data in a hybrid analysis (e.g., \citep{vieleUseHistoricalControl2014, liRevisitTestThen2020}). Others aim to minimize the mean squared error of a combined RCT-external data estimator (e.g., \citep{yangElasticIntegrativeAnalysis2023, chenMinimaxRatesAdaptivity2021,  chengAdaptiveCombinationRandomized2021, oberstBiasrobustIntegrationObservational2022}), with various criteria for including external data or for defining optimal weighted combinations of RCT and external data. Compared to pooling alone, these methods decrease bias but also have lower potential power gains. Compared to analysis of only RCT data, these methods have the potential to increase power but also may increase type 1 error or mean squared error \citep{dejardinUseHistoricalControl2018,  vieleUseHistoricalControl2014, galweySupplementationClinicalTrial2017, cuffeInclusionHistoricalControl2011,  harunCriticalAppraisalBayesian2020,oberstBiasrobustIntegrationObservational2022}. Nonetheless, this intermediate level of evidence provided by data fusion estimators may be valuable when used for underpowered secondary endpoints in rare disease trials. For severe diseases without effective treatments or pediatric approvals of medications that have been shown to be safe and efficacious in adults, inclusion of external control data may allow more trial participants to be randomized to receive a potentially beneficial medication instead of placebo \citep{vieleUseHistoricalControl2014, dejardinUseHistoricalControl2018}. \par
In this manuscript, we explain the challenge of estimating a causal target parameter from a hybrid RCT-external data study using the Causal Roadmap framework \citep{petersenCausalModelsLearning2014a, dang_causal_2023}. Framing the decision  of whether to integrate external data as a problem of data-adaptive experiment selection, we develop a novel cross-validated targeted maximum likelihood estimator for hybrid randomized-external data studies, the ``ES-CVTMLE''. This estimator introduces a TMLE option to the arsenal of data fusion methods. ES-CVTMLE uses cross-validation to separate bias estimation from effect estimation, an approach that has been shown to decrease finite sample bias for other data-adaptive target parameters \citep{montoya_estimators_2023, van_der_laan_targeted_2015}. ES-CVTMLE constructs confidence intervals by sampling from the estimated limit distribution of this estimator, where the sampling process includes an estimate of the bias, further promoting accurate inference. Finally, we consider how to use an estimate of the average treatment effect on a negative control outcome (NCO) to help distinguish biased from unbiased external data. \par
 The remainder of this paper is organized as follows. In Section \ref{sec2}, we describe a causal model and consider causal estimands for hybrid randomized-external data studies. In Section \ref{sec3} we define assumptions necessary for identification of these causal parameters. In Section \ref{sec4}, we discuss estimation of bias due to inclusion of external data. In Section \ref{sec5}, we introduce potential criteria for including external data based on optimizing the bias-variance tradeoff and utilizing the estimated effect of treatment on an NCO. In Section \ref{sec6}, we develop an extension of the cross-validated targeted maximum likelihood estimator (CV-TMLE), an early example of cross-fit double machine learning \citep{zhengAsymptoticTheoryCrossValidated2010, hubbardStatisticalInferenceData2016}, for this new context of data-adaptive experiment selection and define the limit distribution of this ``ES-CVTMLE'' estimator under varying amounts of bias. In Section \ref{sec7}, we provide a summary of the methods. In Section \ref{sec8}, we set up a simulation to assess the performance of ES-CVTMLE compared to adjusted and unadjusted estimators using RCT-only data, a test-then-pool approach \citep{vieleUseHistoricalControl2014}, one method of Bayesian dynamic borrowing \citep{schmidliRobustMetaanalyticpredictivePriors2014}, and a difference-in-differences (DID) approach to adjusting for bias based on an NCO \citep{soferNegativeOutcomeControl2016, shiSelectiveReviewNegative2020}. In Section \ref{sec9}, we evaluate the ability of ES-CVTMLE to distinguish biased from unbiased external controls in a re-analysis of LEADER trial data (NCT01179048) to estimate the effect of liraglutide on glycemic control for patients with Type 2 Diabetes. 

 \section{Causal Model and Causal Estimand for Hybrid Randomized-External Data Studies} \label{sec2}
 For a hybrid RCT-external data study, let $S$ indicate the study in which an individual participated, where $S=0$ indicates participation in the RCT and $S \in \{1,...,K\}$ indicates participation in one of $K$ observational cohorts. We have a binary intervention, $A$, with $A=1$ indicating the treatment of interest and $A=0$ indicating standard-of-care. In this paper, we focus on augmenting only the control arm of an RCT with external data, a scenario which is relevant when the treatment of interest has yet to be approved. The methods presented readily extend to contexts when active treatment is available in the external data, as well. We denote a set of baseline covariates as $W$, and our outcome as $Y$. \par
\begin{figure}[H]
    \centering
    \includegraphics[height=35mm]{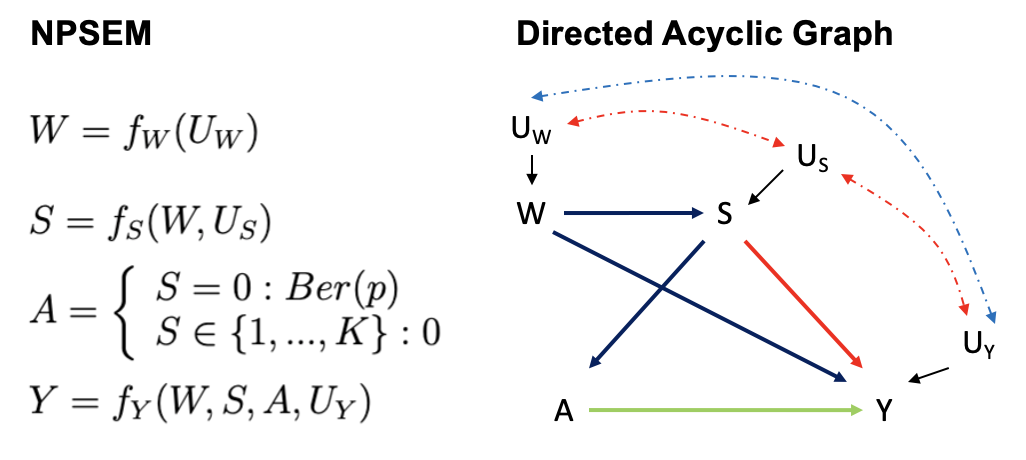}
    \caption{Structural Causal Model}
    \label{fig: 1}
     \footnotesize{\textbf{Caption:} NPSEM: Non-Parametric Structural Equation Model.}
\end{figure}

 As depicted in the non-parametric structural equation model (NPSEM) and directed acyclic graph (DAG) \citep{pearlCausalityModelsReasoning2009} in Figure \ref{fig: 1}, $W$ may affect inclusion in the RCT versus external data. Treatment, $A$, is randomized for those in the RCT and set to $0$ (standard-of-care) for those in the external data, because the treatment has yet to be approved. Thus, $A$ is only affected by $S$ and the randomization probability $p$, not directly by $W$ or any exogenous error. $Y$ may be affected by $W$, $A$, and potentially also directly by $S$ (for example, if trial participation affects quality or type of care beyond access to treatment).  The unmeasured exogenous errors for each of these variables, $U=(U_{W}, U_S, U_Y)$, could potentially be dependent. Our observed data are $n$ independent and identically distributed observations $O_i = (W_{i},S_i,A_i,Y_i)$ with true distribution $P_0$. Supplementary Table \ref{s1} in Appendix \ref{a1} contains a list of symbols used in this manuscript.  \par

A common causal target parameter (estimand) for RCTs is the average treatment effect (ATE). With counterfactual outcomes \citep{neymanApplicationsTheorieProbabilites1923} defined as the outcome an individual would have had if they had received treatment ($Y^{1}$) or if they had received standard of care ($Y^{0}$), the ATE for the RCT participants is defined as 
\begin{center}
    $\Psi^{F}(P_{U,O}) = E_{W|S=0}[E(Y^{1}-Y^{0}|W,S=0)]$.
\end{center}
where, $P_{U,O}$ denotes the joint distribution of $(U,O)$, and $\Psi^F(P_{U,O})$ denotes the target causal parameter. The superscript $F$ indicates that the causal parameter is a function of the distribution of the full data, which includes both observed variables and counterfactual outcomes that cannot be simultaneously observed. 

Next, we consider augmenting the RCT data with external data. Let $X=s$ denote a hybrid experiment in which we augment the RCT with additional controls from external dataset \textit{s}, and let $X=0$ be an experiment that only includes the RCT participants, equivalent to $S=0$. We could alternatively consider hybrid experiments that augment the RCT with potentially more than one additional dataset, or with a weighted combination of RCT and additional data. However, for notational simplicity, we focus here on hybrid designs that incorporate a single external control dataset. $W | X=s$ is then a mixture distribution of $W$ covariates in the RCT and external populations. The probability of selecting a trial or external data participant in this mixture is determined by the proportion of RCT versus external data in the hybrid experiment. Practically, this means that the distribution of baseline covariates, such as age, may be different in the RCT and pooled RCT-external data target populations. However, as long as the external data are restricted to individuals who would have met trial inclusion and exclusion criteria, the target population represented by the distribution of $W | X=s$ is still a population of trial-eligible individuals. The external population may be further restricted to be more similar to the RCT population (e.g., by matching), if desired. 

We then can define multiple potential causal parameters that we could target, one for each potential experiment, where those parameters only differ due to differences in the distribution of $W$ for the different augmented datasets:
\begin{center}
    $\Psi_{s}^{F}(P_{U,O}) = E_{W| X=s}[E(Y^{1}-Y^{0}|W,X=0)]$ for $s \in \{0,...,K\}$.
\end{center}

The index by $s$ denotes a given choice of external data source for augmentation.  

\section{Identification} \label{sec3}

One option for estimating this type of causal parameter using combined RCT and external data would be to make the following assumptions, similar to related assumptions in the causal transport literature (e.g., \citep{rudolphRobustEstimationEncouragement2017, dahabrehGeneralizingCausalInferences2019a}).  \\
\textbf{Assumption 1} (randomization in the trial): $E[Y^{a}|W,X=0,A=a] = E[Y^{a}|W,X=0]$ for each $a \in \{0,1\}$.\\
\textbf{Assumption 2}
(equivalence of covariate-adjusted mean outcomes for trial and combined controls):\\
\begin{center}
    $E_{W|X=s}[E(Y|W,A=0,X=0)] = E_{W|X=s}[E(Y|W,A=0,X=s)].$
\end{center}
 
\textbf{Assumption 3} (positivity): \textit{$P(A=a|W=w, X=s) > 0$ for all $a \in \{0,1\}$ and all $w$ for which $P(W=w, X=s) > 0$.} \par
If we were willing to rely on these assumptions, the average treatment effect (within a target population defined by s) would be identified as a function of the observed data for any choice $s$ of experiment for $s \in \{0,...,K\}$ (i.e., any choice of external data for RCT augmentation) by the following statistical estimand:
\begin{center}
    $\Psi_{s}(P_{0}) = E_{W|X=s}[E(Y|A=1,W,X=0) - E(Y|A=0,W,X=s)]$.
\end{center}

The statistical estimand $\Psi_{s}(P_{0})$ is the difference in the mean outcome associated with treatment versus standard of care, after adjustment for the $W$ covariates, for the specified population defined by experiment $X=s$. In contrast with the causal parameter that is a function of the distribution of counterfactual outcomes, the statistical estimand is only a function of the distribution of the observed data.   Appendix \ref{a2} has a proof of this identification result under \textbf{Assumptions 1-3}. Of note, we do not require an additional assumption that $$E_{W|X=s}[E(Y|W,A=1,X=0)] = E_{W|X=s}[E(Y|W,A=1,X=s)]$$ in this context, although this assumption would also be necessary if treatment $A=1$ were available outside the RCT.

For some choices of $S$, including external data will improve power. However, if \textbf{Assumptions 1, 2, } or \textbf{3} are false, $\Psi_{s}(P_{0})$ does not have a causal interpretation. \textbf{Assumption 1} is true by design. \textbf{Assumption 3} is true in the RCT by design and may be satisfied for other experiments by removing external data controls whose $W$ covariates do not have support in the trial population. The plausibility of \textbf{Assumption 2} requires further consideration.  

\textbf{Assumption 2} is similar to two other assumptions that are commonly discussed in the causal transport literature: $Y \perp S | A, W$ (``S-ignorability'' \citep{bareinboimCausalInferenceDatafusion2016}) and $E[Y^a|W,X=0] = E[Y^a|W]$ (``mean generalizability'' \citep{dahabrehGeneralizingCausalInferences2019a}). Any of these assumptions will be violated if participation in the trial directly affects the outcome beyond assignment to treatment \citep{rudolphRobustEstimationEncouragement2017, dahabrehGeneralizingCausalInferences2019a}. This phenomenon, known as a ``trial effect'', may occur for many reasons including: 1) if patients in the standard of care arm receive better care than patients in the general healthcare system due to closer monitoring or access to the best clinicians, 2) if there is a placebo effect, or 3) if adherence or other behavior changes as a result of trial participation \citep{braunholtz_are_2001, dahabreh_generalizing_2019}. \textbf{Assumption 2} may also be violated if outcomes are measured differently in the trial and the external data. Finally, \textbf{Assumption 2} may be false if there are unmeasured patient-level effect modifiers that differ between the RCT and external study participants \citep{rudolphRobustEstimationEncouragement2017, dahabrehGeneralizingCausalInferences2019a}. For example, if trial participants are generally of a higher socioeconomic status (SES) than participants in the RWD, SES also affects outcomes under standard of care, and SES is not measured, the ATE will not be identified from the combined RCT and external data. Dahabreh et al. \cite{dahabrehGeneralizingCausalInferences2019a}  note that hybrid trials involving pragmatic RCTs integrated with external data from the same healthcare system may be less likely to suffer from these issues. Nonetheless, we may not be certain whether \textbf{Assumption 2} - or the related assumptions of S-ignorability and mean generalizability - are violated in practice. Therefore, we propose an alternative approach that does not rely on this assumption. 

\section{Quantification of the Experiment-Specific Causal Gap} \label{sec4}

Instead, we aim to develop an identification approach relying only on \textbf{Assumptions 1 and 3}, which are known to hold by design in an RCT and by limiting the external data to values of $W$ represented in the trial. From this basis,  we then consider potential approaches for augmenting the RCT data, explicitly acknowledging that any such augmentation has the potential to introduce bias by creating a ``causal gap'', or deviation between the average treatment effect in the target population implied by selection of $s$ and the adjusted measure of association (i.e., statistical target parameter) for this population \citep{gruberTargetedLearningFuture2023}. 
Formally, we denote this bias parameter as  

\begin{center}
    $ \Psi^{\#,F}_{s}(P_{U,O}) =\Psi_{s}(P_0) - \Psi^{F}_{s}(P_{U,O}) $
\end{center}
for each experiment with $X=s$. Under \textbf{Assumption 1} of randomization in the RCT and \textbf{Assumption 3} of positivity applied only to the RCT experiment, the causal gap is identifiable from the observed data with statistical estimand\\
\begin{equation}
     \Psi^{\#}_{s}(P_0) = E_{W|X=s}[E[Y|A=0,W,X=0]] - E_{W|X=s}[E[Y|A=0,W,X=s]].
\end{equation} 

Appendix \ref{a3} has a proof of this identification result. Because randomization and positivity are true in the RCT by design, the causal gap is identified based only on realistic assumptions. \par
The causal gap could result from a combination of different sources of bias, described above as reasons why \textbf{Assumption 2} may be violated. It is possible that individual sources of bias may be present even if the net effect of all sources of bias is zero. Nonetheless, if $\Psi^{\#,F}_{s}(P_{U,O})$ is truly zero, this tells us that the causal ATE is identified from the observed data.\par

Because the causal gap is identifiable, we could bias correct our estimate $\Psi_s(P_0)$ by subtracting $\Psi^{\#}_s(P_0)$. Recall that 

$$ \Psi^{F}_{s}(P_{U,O}) =\Psi_{s}(P_0) - \Psi^{\#,F}_{s}(P_{U,O})$$
$$ = E_{W|X=s}[E(Y|A=1,W,X=0)] - E_{W|X=s}[E(Y|A=0,W,X=s)] - $$
$$ (E_{W|X=s}[E[Y|A=0,W,X=0]] - E_{W|X=s}[E[Y|A=0,W,X=s]])$$
$$ = E_{W|X=s}[E(Y|A=1,W,X=0)] - E_{W|X=s}[E[Y|A=0,W,X=0]]. $$

The issue with this approach is that this estimand conditions on the experiment $X=0$, so we would expect no gain in efficiency compared to estimating the sample ATE from the RCT only \citep{balzer_targeted_2016}. Nonetheless, the information from estimating the causal gap may still be used to determine whether to include external data in the analysis.  \par

\section{Potential Experiment Selection Criteria} \label{sec5}

Although we might expect that the causal gap would not be exactly zero when augmenting an RCT with external data, if the causal gap is small, it may still be preferable to select whichever experiment has the optimal bias-variance tradeoff for estimating the causal ATE. Such an approach of determining combinations of RCT and external data that minimize the estimated mean squared error is taken by \citep{yangElasticIntegrativeAnalysis2023, chengAdaptiveCombinationRandomized2021, chenMinimaxRatesAdaptivity2021, oberstBiasrobustIntegrationObservational2022}. Regardless of the approach, a primary challenge for selecting the truly optimal experiment is that the bias must be estimated from the data. Next, we discuss this challenge and then introduce a novel experiment selector that incorporates bias estimates based on both the primary outcome and a negative control outcome.\par
Ideally, we would like to construct a selector that is equivalent to the ``oracle selector'', which -- given perfect knowledge -- would select the experiment (RCT only or augmented with external data) that minimizes the bias-variance tradeoff:
\begin{center}
    $s^{\star}_{0} = \underset{s}{argmin} \frac{\sigma^{2}_{D^{*}_{\Psi_{s}}}}{n} + (\Psi^{\#}_{s}(P_{0}))^{2}$
\end{center}
where 
\begin{center}
    $D^{*}_{\Psi_{s}}(O) = \frac{I(X=s)}{P_0(X=s)} ((\frac{I(A=1)}{P_{0}(A=1|W,X=s)} - \frac{I(A=0)}{P_{0}(A=0|W,X=s)})(Y-E_0[Y|A,W,X=s])$\\
        $+ E_0[Y|A=1,W,X=s]) - E_0[Y|A=0,W,X=s]) - \Psi_{s}(P_0))$\\
\end{center}
is the efficient influence curve of $\Psi_{s}(P_{0})$. While in practice we do not know which experiment will minimize the bias-variance tradeoff, we nonetheless use this oracle selector to define our statistical estimand of interest: $\Psi_{s^{\star}_{0}}(P_{0})$. \par
The primary challenge is that both the bias and variance terms in $s^{\star}_{0}$ must be estimated. We thus define an empirical bias squared plus variance (\textbf{``b2v'')} selector,
\begin{equation}
   s_{n}^{\star} = \underset{s}{argmin} \frac{\hat{\sigma}^{2}_{D^{*}_{\Psi_{s}}}}{n} + (\hat{\Psi}^{\#}_{s}(P_{n}))^{2}.
\end{equation}
If, for a given experiment with $X=s$, $\Psi^{\#}_{s}(P_0)$ were given and small relative to the standard error of the ATE estimator for that experiment, nominal coverage would be expected for the causal target parameter. If bias were large relative to the standard error of the ATE estimator for the RCT, then the external data would be rejected, and only the RCT would be analyzed. One threat to valid inference using this experiment selection criterion is the case where bias is of the same order as the standard error $\sigma_{D^{*}_{\Psi_{s}}}/\sqrt{n}$, risking decreased coverage \citep{yangElasticIntegrativeAnalysis2023}. Because bias and variance must both be estimated, this magnitude of bias is difficult to detect and exclude. Finite sample variability may lead either to overestimation of bias for unbiased external data (resulting in low power) or underestimation of bias similar in magnitude to $\sigma_{D^{*}_{\Psi_{s}}}/\sqrt{n}$ (resulting in decreased coverage). \par
This challenge exists for any method that bases inclusion of external data on differences in the mean or conditional mean outcome under control for a small RCT control arm versus an external data population. Intuitively, if we are not willing to make \textbf{Assumption 2}, information available in the RCT alone is insufficient to estimate bias from including external data in the analysis precisely enough to guarantee inclusion of extra unbiased controls and exclusion of additional controls that could bias the effect estimate; if the RCT contained this precise information about bias, we would be able to estimate the ATE of $A$ on $Y$ from the RCT precisely enough to not require the external data at all. This challenge suggests that having additional knowledge beyond this information may help the selector distinguish between external data that would introduce varying degrees of bias. 

\subsection{Use of Negative Control Outcomes and Other Tuning Parameters to Modify the Experiment Selector} \label{subsec5.1}

Multiple data fusion estimators introduce a tuning parameter that modifies the external data integration procedure to either make it more or less likely that external data will be included. For example, Cheng and Cai \citep{chengAdaptiveCombinationRandomized2021} determine optimal weights for combining RCT and external data estimators via L1-penalized regression, where the weights are selected to minimize the mean squared error. Out of a concern that basing weights on the estimated MSE may still inappropriately include external data that could lead to biased results, they multiply the bias term by a penalty leading to a more conservative estimator \citep{chengAdaptiveCombinationRandomized2021}. The value of this tuning parameter may be selected to minimize empirical MSE in a large independent validation dataset of pooled RCT and observational data \citep{chengAdaptiveCombinationRandomized2021}. If such a dataset is not available, cross-validation is proposed for tuning parameter selection \citep{chengAdaptiveCombinationRandomized2021} but is not evaluated in simulations.  

For their Bayesian Meta-Analytic-Predictive Priors estimator, Schmidli et al. \citep{schmidliRobustMetaanalyticpredictivePriors2014} estimate $\theta_{\star}$, the mean outcome of controls in the RCT experiment. The prior distribution of $\theta$ is assumed to be Normal$(\mu, \tau^2)$ and is informed by estimates of the mean outcome from external control datasets. $\tau$ is an estimate of the between-study heterogeneity that determines how much external control information is borrowed. A sampling distribution of $\theta$ is generated using a Markov Chain Monte Carlo algorithm and approximated with a mixture of conjugate prior distributions \citep{schmidliRobustMetaanalyticpredictivePriors2014}. To protect against non-exchangeability between external and trial controls, it is recommended to add a unit information prior component to this mixture \citep{schmidliRobustMetaanalyticpredictivePriors2014, weberRBesTBayesianEvidence2021}. The weight of that vague prior must be specified by the researchers based on their beliefs regarding how likely the available control groups are to be exchangeable \citep{schmidliRobustMetaanalyticpredictivePriors2014}. 

These examples show two ways in which data fusion estimators introduce tuning parameters to modify the decision of whether and how to integrate RCT and external data. These methods rely either on additional data of the same type or a subjective decision based on subject-matter knowledge. In contrast, we consider whether there is another source of information that could be estimated from the data (rather than specified subjectively) that could be used to modify the empirical bias-variance tradeoff in the ES-CVTMLE selector. One additional source of information regarding bias is the estimated effect of the treatment on a negative control outcome (NCO). NCOs have previously been used to detect or adjust for bias in observational studies \citep{lipsitchNegativeControlsTool2010, shiSelectiveReviewNegative2020}, but strong, untestable assumptions are required to conclude that the magnitude of bias has been accurately estimated using an NCO. Below, we consider how an NCO may be used to detect bias in the hybrid trial setting while minimizing reliance on new identification assumptions by augmenting our original bias estimate with information from an NCO, rather than relying on the NCO alone. \par
\begin{figure}
    \centering
    \includegraphics[height=3.5cm]{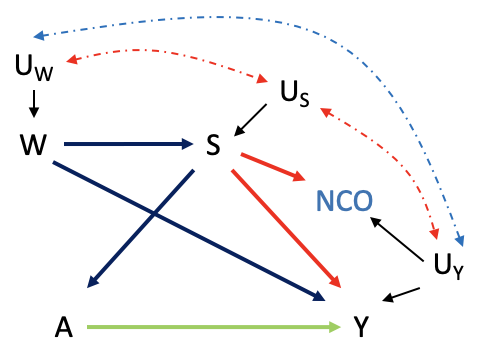}
    \caption{Directed Acyclic Graph Including NCO}
    \label{fig: 2}
         \footnotesize{\textbf{Caption:} NCO: Negative Control Outcome}
\end{figure}

First, we consider how to select an appropriate NCO for a hybrid trial to detect relevant sources of bias. As demonstrated by \citep{lipsitchNegativeControlsTool2010} in the context of confounding bias and \citep{arnoldBriefReportNegative2016} in the context of selection bias and measurement error, an NCO is not affected by treatment but should otherwise share the same causal structure as the true outcome with respect to variables related to potential sources of bias. For example, to evaluate for confounding bias, the NCO should be affected by the same unmeasured factors that are associated with both the treatment and the outcome (known as the ``U-comparability'' assumption) \citep{lipsitchNegativeControlsTool2010}. To detect a basic form of selection bias in which the selection variable is a descendent of both the exposure and the outcome, the NCO should also affect that selection variable \citep{arnoldBriefReportNegative2016}. Extending these principles to the context of augmenting an RCT with external controls, we would aim to choose an NCO with the structure depicted in Figure \ref{fig: 2}; any unmeasured common causes of trial participation and the outcome should also affect the NCO, and if trial participation has a direct effect on outcomes, it should also have a direct effect on the NCO. \par
Once an appropriate NCO has been selected, it may be used to evaluate the causal gap as follows. The statistical estimand for the ATE on the NCO for an experiment with $X=s$ is given by,
$$\Phi_{s}(P_0) = E_{W|X=s}[E[NCO | W, A=1,X=s] - E[NCO | W, A=0, X=s]].$$  
Because treatment does not affect the NCO, a non-zero estimate $\hat{\Phi}_{s}(P_n)$ is either due to finite sample variability or due to factors that lead to a causal gap for the primary outcome under the causal model shown in Figure \ref{fig: 2}. Under an additional set of untestable assumptions, discussed in Appendix \ref{a4}, the causal gap, $\Psi^{\#,F}_{s}(P_{U,O})$, is equivalent to the statistical estimand for the ATE of treatment on the NCO, $\Phi_{s}(P_0)$. Under these assumptions, the statistical estimand $\Psi_{s}(P_0)$ could be bias-corrected by subtracting $\Phi_{s}(P_0)$ \citep{soferNegativeOutcomeControl2016, shiSelectiveReviewNegative2020}. However, because our approach is premised on minimizing additional identification assumptions, we consider a less stringent assumption that allows for detection (rather than correction) of bias using an NCO. \\
\\
\textbf{Assumption 4}: For a given experiment with $X=s$, $\Phi_{s}(P_0) < 0$ \textit{if and only if} $\Psi^{\#,F}_{s}(P_{U,O}) < 0$ and $\Phi_{s}(P_0) > 0$ \textit{if and only if} $\Psi^{\#,F}_{s}(P_{U,O}) > 0$. \\

In the context of our hybrid study, this assumption would mean that if the causal gap is zero (i.e., \textbf{Assumptions 1-3} are true), then the true value of $\Phi_{s}(P_0)$ is also zero. If the causal gap is non-zero then the effect of factors contributing to the causal gap is in the same direction as the effect of these factors on the treatment-NCO relationship. As a more concrete example, if we are primarily concerned that there might be a causal gap because trial participants are more closely monitored, leading to better outcomes, compared to external control participants, then we must pick an NCO that we would also expect to be improved by closer monitoring in the trial. \par
Next, we consider how we may utilize the additional information from estimating the ATE on the NCO even if $\Phi_{s}(P_0)$ is not exactly equivalent to the causal gap $\Psi^{\#,F}_{s}(P_{U,O})$. First, we note that -- for each experiment with $X=s$ -- under \textbf{Assumption 4}, $(\Psi^{\#}_{s}(P_0) + \Phi_{s}(P_0))^2 \geq (\Psi^{\#,F}_{s}(P_{U,O}))^2$ because $\Psi^{\#}_{s}(P_0)$ and $\Phi_{s}(P_0)$ have the same sign. One way to augment our initial estimate of the causal gap with information from the NCO would be to add $\Phi_{s}(P_{0})$ to $\Psi^{\#}_{s}(P_{0})$ using the selector 
\begin{center}
    $s^{\star\star}_{0} = \underset{s}{argmin} \frac{\sigma^{2}_{D^{*}_{\Psi_{s}}}}{n} + (\Psi^{\#}_{s}(P_{0}) + \Phi_{s}(P_{0}))^{2}.$
\end{center}
This selector aims to use the statistical estimand for the ATE of treatment on the NCO to inflate the bias term when biased external controls are considered without inflating the bias term when unbiased external controls are available. This idea is similar to the concept of inflating the bias by a penalty term, as proposed by Cheng and Cai \citep{chengAdaptiveCombinationRandomized2021}, but we estimate the penalty term using information from the NCO. As with other data-fusion estimators that use tuning parameters to modify the criteria for including or weighting external data with RCT data, the ES-CVTMLE using this selector no longer directly aims to minimize the empirical MSE. The goal is instead to use all available information about bias to decide whether to include the external data in the analysis.   

We estimate $s^{\star\star}_{0}$ using the empirical ``\textbf{+nco}'' selector:
\begin{equation}
   s_{n}^{\star\star} = \underset{s}{argmin} \frac{\hat{\sigma}^{2}_{D^{*}_{\Psi_{s}}}}{n} + (\hat{\Psi}^{\#}_{s}(P_{n}) + \hat{\Phi}_{s}(P_{n}))^{2}.
\end{equation}
Using the selector $s_{n}^{\star\star}$ is one of many potential ways that information from an NCO could be used to help determine whether external data should be integrated with RCT data in a hybrid trial. We will evaluate the performance of this proposed selector through simulations below. 
\par
We also consider selector ``\textbf{nco only}'' based only on $\hat{\Phi}_{s}(P_{n})$:
\begin{equation}
    s_{n}^{\star\star\star} = \underset{s}{argmin} \frac{\hat{\sigma}^{2}_{D^{*}_{\Psi_{s}}}}{n} + (\hat{\Phi}_{s}(P_{n}))^{2}.
\end{equation}
 Under the stringent assumptions discussed in Appendix \ref{a4}, the selector in (4) directly aims to minimize the empirical bias-variance tradeoff because $\Phi_{s}(P_{0})$ is equivalent to the casual gap. Nonetheless, because we cannot learn from the data what percentage of the true bias is accounted for by this estimate, we choose to combine rather than replace our estimate $\hat{\Psi}^{\#}_{s}(P_{n})$ with this information. We will compare these options with the originally-proposed selector
$s_{n}^{\star}$. Table 1 summarizes strengths and limitations of each experiment selector. 

\begin{center}
\begin{table}[H]
\caption{Choices of Experiment Selector}
\centering
\begin{tabular}{p{1cm}p{6cm}p{5cm}p{5cm}}
\textbf{Selector} & \textbf{Notation} & \textbf{Strengths} & \textbf{Limitations}\\
 \hline
 \textbf{b2v} & $s_{n}^{\star} = \underset{s}{argmin} \frac{\hat{\sigma}^{2}_{D^{*}_{\Psi_{s}}}}{n} + (\hat{\Psi}^{\#}_{s}(P_{n}))^{2}$ & Does not require an NCO. & May lead to decreased coverage when bias $\approx$ standard error. \\
 \hline
 \textbf{+nco} & $s_{n}^{\star\star} = \underset{s}{argmin} \frac{\hat{\sigma}^{2}_{D^{*}_{\Psi_{s}}}}{n} + (\hat{\Psi}^{\#}_{s}(P_{n}) + \hat{\Phi}_{s}(P_{n}))^{2}$ & Additional bias estimate based on NCO helps to distinguish biased from unbiased external data. & Requires an appropriate NCO. \\
 \hline
 \textbf{nco only} & $s_{n}^{\star\star\star} = \underset{s}{argmin} \frac{\hat{\sigma}^{2}_{D^{*}_{\Psi_{s}}}}{n} + (\hat{\Phi}_{s}(P_{n}))^{2}$ & Alternate estimate of bias$^2$ + variance.  & Relies on strong assumptions (Appendix \ref{a4}).\\
\end{tabular}
\end{table}
\end{center}
\vspace{-0.8cm}
NCO: Negative control outcome.

\section{CV-TMLE for Data-Adaptive Experiment Selection} \label{sec6}

Now that we have defined potential experiment-selection criteria, we must use the data both to select and analyze the optimal experiment. If we select $s^{\star}_{n}$ in a manner that is not outcome-blind, we should not expect to obtain valid inference if we both select the experiment and evaluate our target parameter based on the same data \citep{hubbardStatisticalInferenceData2016}. Cross-validated targeted maximum likelihood estimation (CV-TMLE) was previously developed as a method to obtain valid inference for other data-adaptive target parameters \citep{zhengAsymptoticTheoryCrossValidated2010, hubbardStatisticalInferenceData2016, van_der_laan_targeted_2015}. For example, past work on estimation of the mean outcome under an optimal dynamic treatment rule has demonstrated reduced finite sample bias using a CV-TMLE compared to non-cross-validated estimators \citep{van_der_laan_targeted_2015, montoya_estimators_2023}. We build on this previous work by developing a CV-TMLE for hybrid randomized-external data studies (ES-CVTMLE), which poses new challenges for inference, described below. \par
\begin{figure}[H]
    \centering
    \includegraphics[height=80mm]{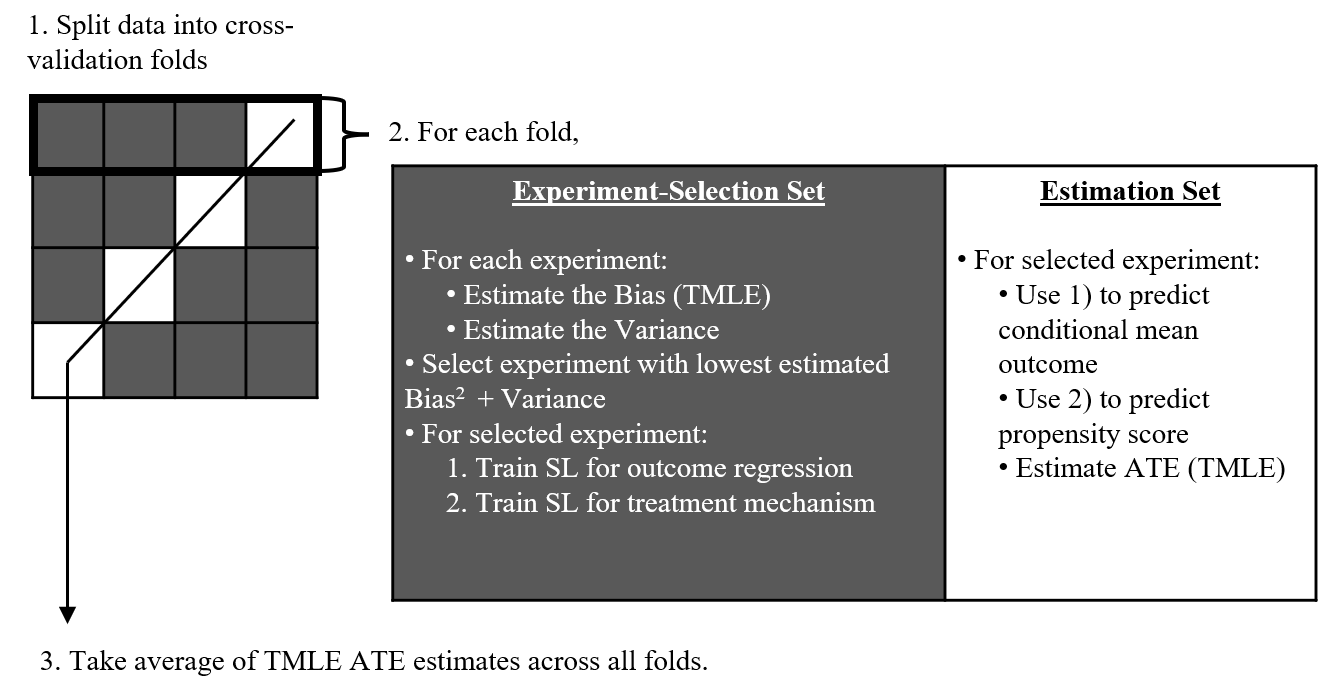}
    \captionof{figure}{Steps for obtaining a point estimate for the ES-CVTMLE target parameter.}
    \label{fig: 3}
     \footnotesize{\textbf{Caption:} TMLE: Targeted Maximum Likelihood Estimation. SL: Super Learner. ATE: Average Treatment Effect.}
    \end{figure}

To understand how the ES-CVTMLE process works, we will use the depiction in Figure \ref{fig: 3} and consider a simple scenario where there are only two experiments: RCT only or RCT-augmented with one external dataset. We start by considering the case where we are using the simple empirical bias-variance tradeoff selector, $s_n^{\star}$ in (2) above. We first split the data into $V$ cross-validation folds. For each fold, we use the experiment-selection data to estimate the bias, $\hat{\Psi}_s^{\#}(P_n)$, for the hybrid RCT-external data experiment using TMLE. $\hat{\Psi}_s^{\#}(P_n)$ for the RCT experiment is 0. The TMLE procedure for estimating the bias is described in greater detail in Appendix E. TMLE is a doubly-robust plug-in estimator that targets initial model fits to optimize the bias-variance tradeoff for the target parameter and that utilizes machine learning to avoid parametric modelling assumptions \citep{vanderlaanTargetedMaximumLikelihood2006a, vanderlaanTargetedLearning2011}. In the case of the ATE, TMLE is asymptotically unbiased if either the outcome regression or the treatment mechanism is estimated consistently and is asymptotically efficient if both are estimated consistently \citep{vanderlaanTargetedLearning2011}. \par
For each fold and each experiment, we also estimate the variance of the ATE estimator, using influence curve-based variance estimates as described in Section 5. We then choose the experiment (RCT alone, or RCT with external data) that has the lowest estimated squared bias plus variance. For each fold, we then estimate the ATE for the selected experiment, as follows.

We use the experiment-selection set data from the selected experiment and the Super Learner ensemble machine learning prediction algorithm \cite{polleySuperLearner2021} to estimate the outcome regression and treatment mechanism/propensity score. If we had selected the hybrid experiment, these would be $E[Y|A,W,X=s]$ and $P(A=1|W,X=s)$, respectively. We use these SL algorithms trained using experiment-selection set data to predict the conditional mean outcome and propensity score for each validation set observation in the selected experiment. We then use these initial validation set estimates in a TMLE to estimate the ATE for the selected experiment. The final target parameter is the average of these TMLE ATE estimates, across all validation folds. An algorithm detailing estimation of the point estimate may be found in Appendix F.

  \par

Next, we introduce the formal notation to describe this process including the possibilities that we may consider multiple potential experiments and any of the three proposed selectors. Randomly splitting the data into $V$ folds yields an experiment-selection set consisting of $(V-1)/V$ of the data and an estimation set consisting of $1/V$. For each split, $v$, the estimation set has empirical distribution $P_{n,v}$ with estimation set subjects assigned $\bar{V}_i=v$. The experiment-selection set has empirical distribution $P_{n,v^{c}}$, and therefore the experiment-selection observations have $\bar{V}_i\neq v$. For each split, the experiment-selection set is used to estimate the causal gap and select the optimal experiment using selector $s^{\star}_{n}(v^{c})$ or $s^{\star\star}_{n}(v^{c})$. The fold-specific target parameter then becomes $\Psi^{F}_{s^{\star}_{n}(v^{c})}(P_{U,O})$, the causal ATE of A on Y in the experiment selected based on the experiment-selection set for fold $v$. The overall target parameter, $\psi_{0}$, and statistical estimand, $\psi_{n,0}$, are then averages of the fold-specific parameters and estimands
\begin{center}
    $\psi_{0} = \frac{1}{V}\sum_{v=1}^{V}\Psi^{F}_{s^{\star}_{n}(v^{c})}(P_{U,O})$\\
    $\psi_{n,0} = \frac{1}{V}\sum_{v=1}^{V}\Psi_{s^{\star}_{n}(v^{c})}(P_{0}).$
\end{center}
 $\psi_{0}$ may be interpreted as the RCT average treatment effect for a target population that is a weighted average of the RCT and external control populations. Because the external control participants must not have any values of the adjustment variables that are not represented in the RCT, we can interpret $\psi_{0}$ more broadly as the ATE in a population consistent with the target population of the RCT. Our modified ES-CVTMLE estimator for data-adaptive experiment-selection is then
\begin{center}
    $\psi_{n} = \frac{1}{V}\sum_{v=1}^{V}\hat{\Psi}^{TMLE}_{s^{\star}_{n}(v^{c})}(P_{n,v})$
\end{center}

\subsection{Asymptotic Distribution of ES-CVTMLE} \label{subsec6.1}

Next, we examine the asymptotic distribution of ES-CVTMLE. Unlike the CV-TMLE for data-adaptive target parameter (DATP) estimation \citep{zhengAsymptoticTheoryCrossValidated2010}, the limit distribution of ES-CVTMLE depends on the amount of bias introduced by a given external dataset. The finite sample challenge for selecting an optimal experiment depends on the magnitude of this true bias relative to the standard error of the ATE estimator, which in turn depends on the sample size. Leeb and Potscher \citep{leeb_model_2005} discuss the challenge of estimating the distribution of a DATP that depends on an unknown parameter, even if that parameter may be estimated. They show that it is generally impossible to obtain a finite sample estimator of the distribution of such a DATP that converges uniformly to the limit distribution. Nonetheless, they assert that ``an asymptotic analysis based on the fiction of a true parameter that depends on sample size provides highly accurate insights into the finite-sample properties of such estimators'' \citep{leeb_model_2005}. We adopt this approach (also taken by Yang et al. \citep{yangElasticIntegrativeAnalysis2023} for their elastic integrative analysis estimator), by analyzing the behavior of the selector when the bias is not fixed but rather dependent on the sample size. To accomplish this goal, define $P_{0,n}$ as the true data distribution dependent on \textit{n}. In order to define the limit distribution, Table 2 defines notation for the scaled estimation error for each of the estimated parameters as well as vectors of these scaled estimation errors across experiments and folds.

\begin{table}[H]
\centering
\begin{tabularx}{\textwidth}{>{\raggedright\arraybackslash}X|>{\raggedright\arraybackslash}X}

\textbf{Definition} & \textbf{Explanation} \\
\hline
\parbox[t]{\hsize}{\strut $Z_{n}(s,v) = \sqrt{n}(\hat{\Psi}_s^{TMLE}(P_{n,v}) - \Psi_{s}(P_0))$ \\ $\approx \frac{1}{\sqrt{n}}\sum_{i=1}^{n}D^{*}_{\Psi_{s},v}(O_i,V_i)$\strut} & $Z_{n}(s,v)$ is the scaled estimation error for the TMLE ATE estimator for experiment $s$ and fold $v$. \\
\hline
\parbox[t]{\hsize}{\strut $Z_{n}=(Z_{n}(s,v):s=0,\ldots,K,v=1,\ldots,V) \overset{D}{\rightarrow}$ \\ $Z=(Z(s,v):s=0,\ldots,K,v=1,\ldots,V)$\strut} & $Z_n$ is a vector of these $Z_n(s,v)$ across all experiments and folds, which converges in distribution to the vector $Z$. \\
\hline
\parbox[t]{\hsize}{\strut $Z^{\#}_{n}(s,v) = \sqrt{n}(\hat{\Psi}^{\#,TMLE}_{s}(P_{n,v^c}) - \Psi^{\#}_{s}(P_{0,n})) \approx \frac{1}{\sqrt{n}}\sum_{i=1}^{n}D^{*}_{\Psi^{\#}_{s},v^c}(O_i,V_i)$\strut} & $Z^{\#}_{n}(s,v)$ is the scaled estimation error for the TMLE causal gap estimator for experiment $s$ and fold $v$. \\
\hline
\parbox[t]{\hsize}{\strut $Z^{\#}_{n}=(Z^{\#}_{n}(s,v):s,v) \overset{D}{\rightarrow}$ \\ $Z^{\#}=(Z^{\#}(s,v):s,v)$\strut} & $Z^{\#}_{n}$ is a vector of these $Z^{\#}_n(s,v)$ across all experiments and folds, which converges in distribution to the vector $Z^{\#}$. \\
\hline
\parbox[t]{\hsize}{\strut $Z^{\#+\Phi}_{n}(s,v) = \sqrt{n}(\hat{\Psi}^{\#,TMLE}_{s}(P_{n,v^c}) + \hat{\Phi}^{TMLE}_{s}(P_{n,v^c})$ \\ $ - (\Psi^{\#}_{s}(P_{0,n}) + \Phi_{s}(P_{0,n}))) \approx \frac{1}{\sqrt{n}}\sum_{i=1}^{n}(D^{*}_{\Psi^{\#}_{s},v^c}(O_i,V_i) + D^{*}_{\Phi_{s},v^c}(O_i,V_i))$\strut} & $Z^{\#+\Phi}_{n}(s,v)$ is the scaled estimation error for the TMLE causal gap estimator + ATE on NCO estimator for experiment $s$ and fold $v$. \\
\hline
\parbox[t]{\hsize}{\strut $Z^{\#+\Phi}_{n}=(Z^{\#+\Phi}_{n}(s,v):s,v) \overset{D}{\rightarrow}$ \\ $Z^{\#+\Phi}=(Z^{\#+\Phi}(s,v):s,v)$\strut} & $Z^{\#+\Phi}_{n}$ is a vector of these $Z^{\#+\Phi}_{n}(s,v)$ across all experiments and folds, which converges in distribution to the vector $Z^{\#+\Phi}$. \\
\end{tabularx}
\caption{Definitions relevant for understanding the asymptotic distribution of ES-CVTMLE}
\label{tab:wrapped_text}
\end{table}

We also define the influence curves of the parameters that we will estimate. The influence curve for an observation can be thought of as the influence of that observation on the value of the target parameter, and the influence curve evaluated at all observations in a dataset may be used to obtain variance and confidence interval estimates \citep{hampel_influence_1974}. \par 

$$D^{*}_{\Psi_{s},v}(O,\bar{V}) = \frac{I(X=s,\bar{V}=v)}{P(X=s,\bar{V}=v)}((\frac{I(A=1)}{P(A=1|W,X=s)} - \frac{I(A=0)}{P(A=0|W,X=s)})(Y-E[Y|A,W,X=s]) $$
   $$ + E[Y|A=1,W,X=s] - E[Y|A=0,W,X=s] - \Psi_{s}(P_0))$$

is the efficient influence curve of the ATE for experiment $s$ and fold $v$. 

 $$D^{*}_{\Psi^{\#}_{s},v^c}(O,\bar{V}) = \frac{I(X=s,\bar{V} \neq v)}{P(X=s,\bar{V} \neq v)} ((\frac{I(S=0,A=0)}{P(S=0,A=0|W,X=s)})(Y - E[Y|A,W,S,X=s])$$
    $$-\frac{I(A=0)}{P(A=0|W,X=s)}(Y - E[Y|A,W,X=s] + E[Y|A=0,W,S=0,X=s] - E[Y|A=0,W,X=s] - \Psi^{\#}_{s}(P_{0,n}))$$
is the efficient influence curve of the causal gap parameter for experiment $s$ and fold $v$. 

 $$D^{*}_{\Phi_{s},v^c}(O,\bar{V}) = \frac{I(X=s,\bar{V}\neq v)}{P(X=s,\bar{V}\neq v)}((\frac{I(A=1)}{P(A=1|W,X=s)} - \frac{I(A=0)}{P(A=0|W,X=s)})(NCO-E[NCO|A,W,X=s])$$
 $$+E[NCO|A=1,W,X=s] - E[NCO|A=0,W,X=s] - \Phi_s(P_{0,n}))$$

is the efficient influence curve for the ATE of treatment on the NCO for experiment $s$ and fold $v$, and 
$$D^{*}_{(\#+\Phi)_{s},v^c}(O,\bar{V})= D^{*}_{\Psi^{\#}_{s},v^c}(O,\bar{V}) + D^{*}_{\Phi_{s},v^c}(O,\bar{V}).$$

Next, we consider the distribution of the selectors, which are random variables that depend on the distribution of $Z^{\#}_{n}(s,v)$ or $Z^{\# + \Phi}_{n}(s,v)$: 

\begin{center}
    $s_{n}^{\star}(v^c)  = \underset{s}{argmin} \hspace{0.1cm} \hat{\sigma}^{2}_{D^{*}_{\Psi_{s,n,v^c}}}  + (Z^{\#}_{n}(s,v) + \sqrt{n}(\Psi^{\#}_{s}(P_{0,n})))^{2} $\\
 $s_{n}^{\star\star}(v^c)  = \underset{s}{argmin} \hspace{0.1cm} \hat{\sigma}^{2}_{D^*_{\Psi_{s,n,v^c}}}  + (Z^{\# + \Phi}_{n}(s,v) + \sqrt{n}(\Psi^{\#}_{s}(P_{0,n}) + \Phi_{s}(P_{0,n})))^{2} $\\
\end{center}
Let $s_{n}^{\star}=(s_{n}^{\star}(v^c):v)$ and $s_{n}^{\star\star}=(s_{n}^{\star\star}(v^c):v)$ represent the multivariate selectors applied across all experiment-selection sets. This means that the distribution of the ES-CVTMLE depends on which experiment is selected in each fold, and the experiment that is selected in each fold depends both on the true magnitude of the causal gap for that experiment and the estimation error for the causal gap. After defining these quantities, we can now consider the limit distribution of the ES-CVTMLE, itself. \par
\bigskip
\textbf{Theorem 1:} Convergence of ES-CVTMLE to an Average of Mixtures of Normal Distributions:\\
\textit{Under conditions of convergence of second-order remainders, consistency of efficient influence curve estimation, and a Donsker class condition for bias term estimation specified in Appendix \ref{a7}},
$s_{n}^{\star}(v^c)$ and $s_{n}^{\star\star}(v^c)$ approximate the limit processes $\bar{S}^{\star}(v^c)$ and $\bar{S}^{\star\star}(v^c)$ such that 
\begin{center}
    $\bar{S}^{\star}(v^c)  \sim \underset{s}{argmin} \hspace{0.1cm} \sigma^{2}_{D^{*}_{\Psi_{s},v^c}}  + (Z^{\#}(s,v) + \sqrt{n}\Psi^{\#}_{s}(P_{0,n}))^{2} $\\
  $\bar{S}^{\star\star}(v^c)  \sim \underset{s}{argmin} \hspace{0.1cm} \sigma^{2}_{D^{*}_{\Psi_{s},v^c}}  + (Z^{\#+\Phi}(s,v) + \sqrt{n}(\Psi^{\#}_{s}(P_{0,n}) + \Phi_{s}(P_{0,n})))^{2} $\\
\end{center}
and the standardized ES-CVTMLE,
\begin{center}
    $ \sqrt{n}(\psi_{n} - \psi_{n,0}) = \frac{1}{V}\sum_{v=1}^{V}(Z(\bar{S}^{\star}(v^c),v)) + o_P(1)$\\
    or $ \sqrt{n}(\psi_{n} - \psi_{n,0}) = \frac{1}{V}\sum_{v=1}^{V}(Z(\bar{S}^{\star\star}(v^c),v)) + o_P(1)$
\end{center}
     converges to an average across sample splits of mixtures of normal distributions. The Proof of Theorem 1 may be found in Appendix \ref{a7}. Intuitively, this makes sense based on the process by which the ES-CVTMLE point estimate is calculated: first, an experiment is selected for each fold, then the ATE for that experiment is estimated, and finally the average of these ATEs across all folds is calculated.

       \begin{figure}[H]
    \centering
    \includegraphics[height=160mm]{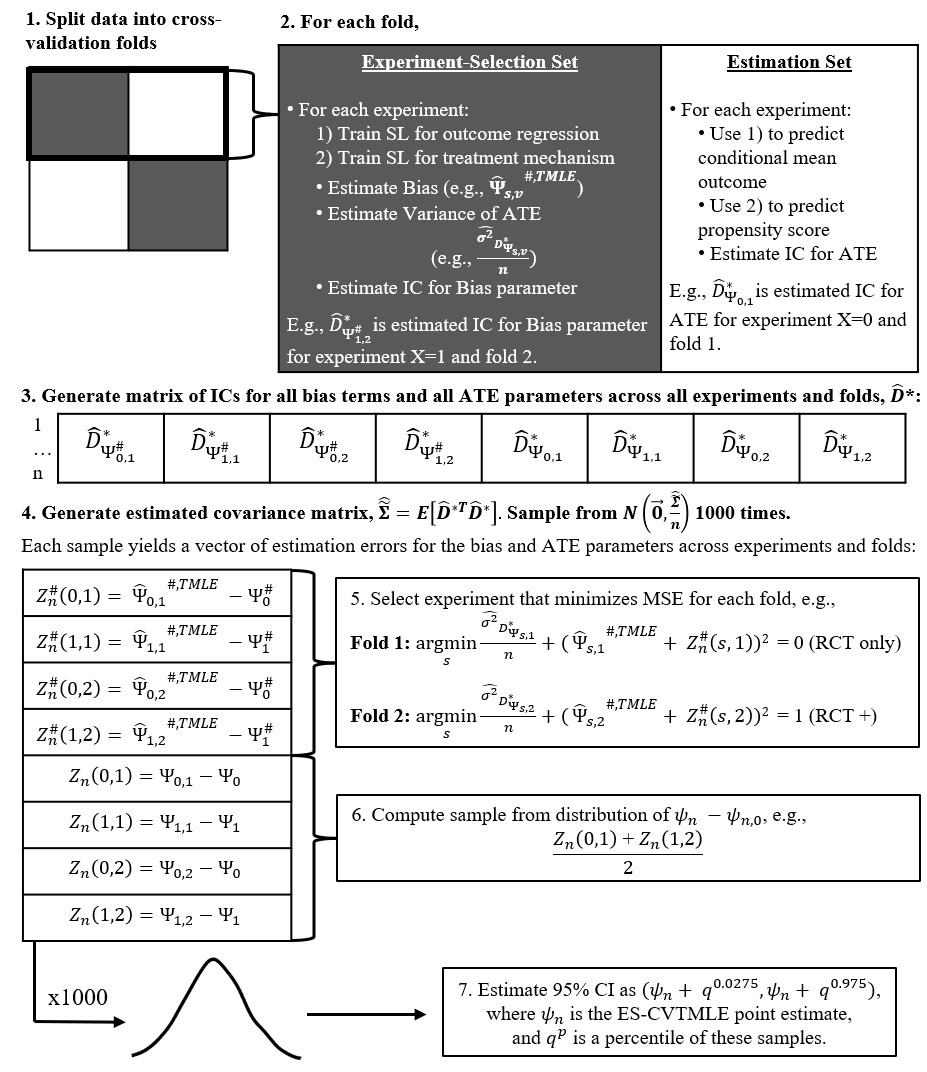}
    \captionof{figure}{Steps for obtaining confidence intervals for the ES-CVTMLE target parameter.}
    \label{fig: 4}
     \footnotesize{\textbf{Caption:} TMLE: Targeted maximum likelihood estimation. SL: Super Learner. IC: Influence Curve. ATE: Average Treatment Effect. MSE: Mean Squared Error. RCT+: Randomized Controlled Trial + external data}.
    \end{figure}\par 

This distribution is defined by a sampling process. Figure \ref{fig: 4} shows a simple example of how to generate confidence intervals for the ES-CVTMLE based on this sampling process when we only have one external dataset and two-fold cross-validation. Appendix \ref{a6} provides a technical algorithm detailing the same steps. Recall that $Z^{\#}$ is a vector of the estimation error for the causal gap parameter across all experiments and folds and $Z$ is a vector of the estimation error for the ATE parameter across all experiments and folds. We will generate samples both of $Z^{\#}$ and $Z$ to incorporate the variance from bias estimation and effect estimation into the procedure for confidence interval construction. As shown in the Proof of Theorem 1, the stacked vector  $\tilde{Z} = (Z^{\#},Z) \sim N(\overset{\rightarrow}{0},\tilde{\Sigma})$, so we must estimate this covariance matrix, $\tilde{\Sigma}$.
     
     Recall that we will use the influence curves of our parameters for variance estimation. To do that, we define a matrix of influence curves, $D^*$ with $n$ rows and $(K+1)*V*2$ columns corresponding to the vector $\tilde{Z}$. For example, if there are two possible experiments and two cross-validation folds, this matrix will have eight columns. If the $i^{th}$ element of $\tilde{Z}$ is the estimation error for the causal gap parameter for experiment $X=1$ and fold 2, then the $i^{th}$ column of the matrix $D^*$ is the influence curve for the causal gap parameter for experiment $X=1$ and fold 2, $D^{*}_{\Psi^{\#}_{1,2}}(O,\bar{V})$, evaluated across all observations in the dataset. For the ES-CVTMLE point estimate, we have already estimated the casual gap and variance for each experiment and each fold as well as the outcome and treatment mechanism regressions necessary to estimate $D^*$. Then, as shown in the Proof of Theorem 1 found in Appendix \ref{a7}, $\tilde{\Sigma} = E[D^{*T}D^{*}]$. If we are incorporating information from the NCO in the selector, then we simply need to substitute $Z^{\# + \Phi}$ for $Z^{\#}$ and $D^{*}_{{(\# + \Phi)}_{1,2}}(O,\bar{V})$ for $D^{*}_{\Psi^{\#}_{1,2}}(O,\bar{V})$ in the process described above.  

  We use Monte Carlo simulation to generate confidence intervals for the ES-CVTMLE as follows. We estimate $\hat{D}^*$ and $\hat{\tilde{\Sigma}}$. We sample from $N(\overset{\rightarrow}{0},\frac{\hat{\tilde{\Sigma}}}{n})$ to generate 1000 samples of $\tilde{Z}$. For each sample, we calculate $\bar{s}^{\star}$ or $\bar{s}^{\star\star}$ and $\frac{1}{V}\sum_{v=1}^{V}(Z(\bar{s}^{\star}(v^c),v))$ or $\frac{1}{V}\sum_{v=1}^{V}(Z(\bar{s}^{\star\star}(v^c),v))$. Finally, we define $95\%$ confidence intervals based on the percentiles $q^p$ of these samples as $(\psi_{n} + q^{0.025}, \psi_{n} + q^{0.975})$. 
  
  In order to implement this process, the last factor to consider is how to calculate $\bar{s}^{\star}$ or $\bar{s}^{\star\star}$ if the selector depends on the unknown bias. Table 3 describes the behavior of this selector when the true magnitude of the causal gap depends on the sample size, considering scenarios in which $\sqrt{n}\Psi^{\#}_{s}(P_{0,n})$ converges in probability to 0 (small bias), a constant (intermediate bias), or infinity (large bias). As shown in Table 3, although the random selector depends on $\Psi^{\#}(P_{0,n})$, it converges to a limit distribution that does not depend on $n$, and which is known if bias is small, known up to a constant if bias is intermediate, and degenerate, selecting 0 with probability 1, if bias is large. In the case where the RCT-only experiment is selected across all folds (large bias), ES-CVTMLE is equivalent to a standard CV-TMLE for the RCT alone, and similarly converges to a normal distribution. In this case, we use influence curve-based variance estimates, consistent with a standard CV-TMLE \citep{zhengAsymptoticTheoryCrossValidated2010}. When some external data are included, ES-CVTMLE converges to an average of mixtures of normal distributions which depends on the unknown bias, if bias is non-negligible. \par
  
  In this case, we may approximate the limit distribution of the selector by plugging in an estimate of the causal gap for the unknown constant in the selector, as shown in Step 5 of Figure \ref{fig: 4}. These TMLE-based plug-in estimators for the bias terms are consistent estimators of the true bias under mild conditions described in the proof of Theorem 1 in Appendix \ref{a7}. Nonetheless, we cannot guarantee $95\%$ confidence interval coverage when we use a plug-in estimate of the bias in our Monte Carlo simulation. If the bias is truly zero, our finite sample estimate will be an over-estimate of the true bias, leading to conservative inference. If the bias is of the same order as the standard error, we may over- or under-estimate the bias in a given finite sample. Yet this finite sample estimate is our best guess of the true bias using the available data. This approach allows us to provide a good approximation of the finite sample distribution of the estimator when bias is non-negligible, as we will show in upcoming simulations.

\begin{center}
\begin{table}[H]
\caption{Limit Distribution of Selector and Estimator with Different Magnitudes of True Bias}
    \centering
    %\begin{tabular}{|p{0.25\linewidth} | p{0.4\linewidth}| p{0.3\linewidth}|}
    \begin{tabular}{p{2.7cm}p{6cm}p{6cm}}
\textbf{Magnitude of Bias} & \textbf{Limit Distribution of Selector} & \textbf{Limit Distribution of Estimator}\\
 \hline
 \textbf{Small}\\
 $\sqrt{n}\Psi^{\#}_{s}(P_{0,n}) \overset{p}{\rightarrow} 0$ &  $\bar{S}^{\star}(v^c)  \sim \underset{s}{argmin} \hspace{0.1cm} \sigma^{2}_{D^{*}_{\Psi_{s},v^c}}  + (Z^{\#}(s,v))^{2} $ & Average of mixtures of normal distributions\\
 \hline 
 \textbf{Intermediate}\\
$\sqrt{n}\Psi^{\#}_{s}(P_{0,n}) \overset{p}{\rightarrow} C^{\dagger}$& $\bar{S}^{\star}(v^c)  \sim \underset{s}{argmin} \hspace{0.1cm} \sigma^{2}_{D^{*}_{\Psi_{s},v^c}}  + (Z^{\#}(s,v) + C)^{2} $ & Average of mixtures of normal distributions depending on unknown C\\
 \hline
\textbf{Large}\\$\sqrt{n}\Psi^{\#}_{s}(P_{0,n}) \overset{p}{\rightarrow} \infty$ & $\bar{S}^{\star}(v^c) = 0$ & Normal distribution\\
\end{tabular}
\end{table}
\end{center}
\vspace{-1.5cm}
$^{\dagger}$C is a constant.\\

\section{Summary of Methods} \label{sec7}

In summary, \textbf{Assumption 2} is a strong untestable assumption that would need to be true for the causal estimand to be identified from the pooled RCT and external data. Violations of \textbf{Assumption 2} will lead to a non-zero bias, and we can estimate this ``causal gap'' from the observed data. Similar to other data fusion estimators \citep{yangElasticIntegrativeAnalysis2023, chengAdaptiveCombinationRandomized2021, chenMinimaxRatesAdaptivity2021, oberstBiasrobustIntegrationObservational2022}, the aim of the ES-CVTMLE is to integrate external with RCT data only if doing so will improve the bias-variance tradeoff for estimating the causal target parameter. We expect good coverage for the causal estimand from an analysis using a) the pooled data if bias due to violations of \textbf{Assumption 2} is negligible relative to the standard error or b) the RCT alone if bias is large and the external data are rejected. The challenge is that the bias must be estimated. Practically, this means that it is difficult to distinguish external data with negligible bias from external data with bias of a magnitude that could impact coverage for the causal estimand. We consider how an NCO may be used to help distinguish biased from unbiased external data, although the challenge of finite sample variability in bias estimates remains. Recognizing that external data with non-negligible bias may be included due to the imprecision of bias estimates in finite samples, our procedure for confidence interval construction assumes that the bias remains non-negligible asymptotically. The result is a more conservative method of constructing confidence intervals than one that assumes that the true bias converges to either zero or infinity asymptotically. The goal of this method is to provide relatively robust inference for the causal estimand even when external data with non-negligible bias are integrated. We evaluate how well the ES-CVTMLE achieves this goal through simulations below.

\section{Simulation} \label{sec8}

Next, we evaluate the performance of ES-CVTMLE compared to an RCT-only unadjusted estimator, an RCT-only CV-TMLE using the \textit{tmle R} package \citep{gruberTmlePackageTargeted2012}, and three other data fusion methods described below across three magnitudes of simulated external data bias.

\subsection{Data Generation} \label{subsec8.1}

We generate a small RCT ($S=0$) of 150 observations with probability of randomization to $A=1$ of 0.67. The goal is to mimic a situation where, for ethical reasons, it is desirable to randomize more participants to active treatment. We also simulate three candidate external datasets $S \in \{1,2,3\}$ of 500 observations each, all with $A=0$. Thus, no treatment is available outside the trial. Dataset $S=1$ has the same data-generating distribution as the RCT except that all $A=0$, so any apparent bias in $S=1$ is due to finite sample variability. There are two unmeasured bias variables $B_{1}$ and $B_{2}$ that are deterministically 0 in $S=0$ and $S=1$ and are generated as follows in $S \in \{2,3\}$. Biased external data could be included if the bias is approximately the square root of the difference between the variance of the ATE estimator for the RCT-only experiment and the variance of the ATE estimator for the hybrid experiment, which is $B \approx 0.21$ in this simulation. We then generate $B_{1}$ and $B_{2}$ as normally distributed random variables such that average total bias in $S=2$ is $\approx B$ (intermediate bias) and in $S=3$ is $\approx 5*B$ (large bias). The outcome, $Y$ is a function of both $B_1$ and $B_2$, while the NCO is only a function of $B_1$, so \textbf{Assumption 4} is true, but the NCO is not affected by all sources of bias (an imperfect NCO). Appendix \ref{a8} contains further details regarding the data generating process and specifications for TMLE-based estimators used in this simulation.

\subsection{Comparators} \label{subsec8.2}

For each combination of $S=0$ with one of $S \in \{1,2,3\}$, we compare ES-CVTMLE with potential selectors $s_{n}^{\star}$ (\textbf{b2v}), $s_{n}^{\star\star}$ (\textbf{+nco}), and $s_{n}^{\star\star\star}$ (\textbf{nco only}) to three other data fusion estimators. These comparators were selected because they were developed for the context of augmenting a control arm of an RCT with external control data, they are commonly referenced, and they include methods for confidence interval construction. We modify the test-then-pool \citep{vieleUseHistoricalControl2014} and NCO-based difference-in-differences \citep{soferNegativeOutcomeControl2016, shiSelectiveReviewNegative2020} approaches by using CV-TMLE estimators of the relevant parameters. \par

\subsubsection{Test-then-Pool} \label{subsubsec8.2.1}

For the ``test-then-pool'' approach \citep{vieleUseHistoricalControl2014}, a hypothesis test is conducted for a difference in the mean outcome of the trial controls and the mean outcome of the external controls. RCT and external data are combined if the null hypothesis is not rejected; if the null hypothesis is rejected, then the RCT data are analyzed without integration of external data \citep{vieleUseHistoricalControl2014}. The original test-then-pool used an unadjusted estimator of the difference in mean outcome under treatment and control \citep{vieleUseHistoricalControl2014}. For the unadjusted version, the hypothesis test for including external data is obtained using Welch's t-test with unequal variances. For the sake of comparison with other TMLE-based estimators, we include an adjusted version that uses CV-TMLE to estimate the ATE of S on Y among those with A=0 and decides to pool RCT and RWD if the $95\%$ confidence interval for this estimate includes zero. We then obtain an estimate of the ATE of A on Y in the pooled or RCT-only sample, again using CV-TMLE. While the ``test-then-pool'' approach has been criticized for inappropriately including biased data due to low power of the test \citep{liRevisitTestThen2020}, a byproduct of this limitation is that the estimator is able to achieve large power gains when unbiased external controls are available. It is thus an interesting comparator as a high-risk, high-reward strategy for data fusion.

\subsubsection{Meta-Analytic-Predictive Priors} \label{subsubsec8.2.2}

For comparison to a method of Bayesian Dynamic Borrowing, we use the \textit{RBesT} \textit{R} package \citep{weberRBesTBayesianEvidence2021} based on \citep{schmidliRobustMetaanalyticpredictivePriors2014}. As described by Schmidli et al. \citep{schmidliRobustMetaanalyticpredictivePriors2014} but modified to avoid notational conflict with this manuscript, $\theta_{\star}$ is the mean outcome of controls in the RCT, and $\theta_s$ is the mean outcome for controls in external dataset $s$. It is then assumed that these parameters come from the same prior distribution, which is Normal$(\mu, \tau^2)$. $\tau$ describes the heterogeneity in $\theta$ across datasets. For a continuous outcome, Weber et al. recommend a Half-Normal$(0, \frac{\sigma}{2})$ prior distribution for $\tau$, where $\sigma$ is the standard deviation of the outcome estimated from external studies \citep{weberRBesTBayesianEvidence2021}. Because the choice of the prior distribution of $\tau$ can impact results, Schmidli et al. recommend sensitivity analyses with different parameterizations of this distribution  \citep{schmidliRobustMetaanalyticpredictivePriors2014}. As described in Section 5, the researcher may specify the weight of a vague prior component to the estimated distribution of $\theta$. Weber et al. (2021) suggest a weight of 0.2 \citep{weberRBesTBayesianEvidence2021}, which is used in the simulations below. \par

Using this meta-analytic-predictive prior based on the external data, the final control target parameter is estimated as the mean of the posterior distribution $E(\theta | O^{n})$. The posterior distribution of the treatment target parameter is estimated as a mixture of conjugate distributions based on a weakly informative unit-information prior \citep{weberRBesTBayesianEvidence2021}.

\subsubsection{Negative Control Outcome (Difference-in-Differences Approach)} \label{subsubsec8.2.3}

Because our methods incorporate information from a negative control outcome, we also compare simulation results to a simple bias adjustment approach that is also based on an NCO. Multiple authors have noted that, under the following assumptions, adjustment for bias due to an unmeasured variable that is associated with both treatment (or in this case study participation) and outcomes using an NCO can be accomplished using a difference-in-differences approach \citep{ soferNegativeOutcomeControl2016, shiSelectiveReviewNegative2020}. The first assumption is U-comparability, which states that all of the unmeasured factors that affect the A-Y relationship are the same as the unmeasured factors that affect the A-NCO relationship \citep{lipsitchNegativeControlsTool2010}. The second is ``additive equi-confounding'', which states that the unmeasured confounding has the same effect (on the additive scale) on the primary outcome as on the NCO \citep{soferNegativeOutcomeControl2016, shiSelectiveReviewNegative2020}. Under these assumptions, an estimator for the average treatment effect of A on Y for a given the experiment with $X=s$ may be defined as $\hat{\Psi}^{DID}_{s}(P_{n}) = \hat{\Psi}_{s}(P_{n}) - \hat{\Phi}_{s}(P_{n})$ \citep{soferNegativeOutcomeControl2016, shiSelectiveReviewNegative2020}. For a consistent comparison with the rest of our methods, we use CV-TMLE to estimate both parameters. The efficient influence curve of $\Psi^{DID}_{s}$ is then $D^{*}_{\Psi^{DID}_{s}} = D^{*}_{\Psi_{s}} - D^{*}_{\Phi_{s}}$. 

\subsection{Simulation Results} \label{subsec8.3}

\begin{table}[H]
\begin{center}
\caption{Results of Simulation}
\label{2}
    \begin{tabular}{@{}lcccccc@{}} 
Estimator (external data) & Bias & Variance & Mean Est. Var. & MSE & Coverage & Power \\
  \hline
  RCT unadjusted estimator & 0.005 & 0.206 & 0.219 & 0.206 & 0.96 & 0.24\\
\hline
  RCT CV-TMLE & 0.004 & 0.065 & 0.070 & 0.065 & 0.95 & 0.64\\
\hline
ES-CVTMLE $s_{n}^{\star}$ (\textbf{b2v}) (S=1) & 0.003 & 0.054 & 0.058 & 0.054 & 0.96 & 0.74 \\
ES-CVTMLE $s_{n}^{\star}$ (\textbf{b2v}) (S=2) & -0.026 & 0.065 & 0.061 & 0.065 & 0.95 & 0.71 \\
ES-CVTMLE $s_{n}^{\star}$ (\textbf{b2v}) (S=3) & 0.005 & 0.065 & 0.071 & 0.065 & 0.95 & 0.64 \\
\hline
ES-CVTMLE $s_{n}^{\star\star}$ (\textbf{+nco}) (S=1) & 0.005 & 0.045 & 0.044 & 0.045 & 0.96 & 0.83\\
ES-CVTMLE $s_{n}^{\star\star}$ (\textbf{+nco}) (S=2) & -0.028 & 0.059 & 0.052 & 0.060 & 0.92 & 0.76\\
ES-CVTMLE $s_{n}^{\star\star}$ (\textbf{+nco}) (S=3) & 0.005 & 0.065 & 0.071 & 0.065 & 0.95 & 0.64\\
\hline
ES-CVTMLE $s_{n}^{\star\star\star}$ (\textbf{nco only}) (S=1) & 0.004 & 0.028 & 0.034 & 0.028 & 0.97 & 0.92\\
ES-CVTMLE $s_{n}^{\star\star\star}$ (\textbf{nco only}) (S=2) & -0.152 & 0.036 & 0.038 & 0.059 & 0.87 & 0.95\\
ES-CVTMLE $s_{n}^{\star\star\star}$ (\textbf{nco only}) (S=3) & -0.037 & 0.089 & 0.068 & 0.090 & 0.91 & 0.67\\
\hline
TTP (CV-TMLE) (S=1) & 0.004 & 0.037 & 0.029 & 0.037 & 0.93 & 0.93\\
TTP (CV-TMLE) (S=2) & -0.113 & 0.059 & 0.033 & 0.072 & 0.79 & 0.87\\
TTP (CV-TMLE) (S=3) & 0.004 & 0.065 & 0.070 & 0.065 & 0.95 & 0.64\\
\hline
Diff-in-Diff (NCO) (S=1) & 0.008 & 0.052 & 0.054 & 0.052 & 0.95 & 0.73\\
Diff-in-Diff (NCO) (S=2) & -0.040 & 0.054 & 0.054 & 0.056 & 0.94 & 0.79\\
Diff-in-Diff (NCO) (S=3) & -0.227 & 0.054 & 0.054 & 0.105 & 0.84 & 0.94\\
\hline
TTP (T-Test) (S=1) & -0.001 & 0.122 & 0.090 & 0.122 & 0.93 & 0.53\\
TTP (T-Test) (S=2) & -0.132 & 0.147 & 0.095 & 0.164 & 0.88 & 0.70\\
TTP (T-Test) (S=3) & -0.128 & 0.359 & 0.182 & 0.376 & 0.76 & 0.35\\
\hline
\textit{RBesT} \citep{weberRBesTBayesianEvidence2021} (S=1) & -0.005 & 0.152 & 0.183 & 0.152 & 0.97 & 0.29\\
\textit{RBesT} \citep{weberRBesTBayesianEvidence2021} (S=2) & -0.052 & 0.157 & 0.185 & 0.159 & 0.96 & 0.32 \\
\textit{RBesT} \citep{weberRBesTBayesianEvidence2021} (S=3) & -0.116 & 0.213 & 0.222 & 0.227 & 0.94 & 0.31 \\
\end{tabular}
\end{center} 
Mean. Est. Var.: Mean of variance estimates. S=1: unbiased external data. S=2: external data with intermediate bias. S=3: external data with large bias. Power: Probability that confidence interval $< 0$ across 1000 iterations. TTP: Test-then-Pool. $s_{n}^{\star} = \underset{s}{argmin} \frac{\hat{\sigma}^{2}_{D^{*}_{\Psi_{s}}}}{n} + (\hat{\Psi}^{\#}_{s}(P_{n}))^{2}$, $s_{n}^{\star\star} = \underset{s}{argmin} \frac{\hat{\sigma}^{2}_{D^{*}_{\Psi_{s}}}}{n} + (\hat{\Psi}^{\#}_{s}(P_{n}) + \hat{\Phi}_{s}(P_{n}))^{2}$, $s_{n}^{\star\star\star} = \underset{s}{argmin} \frac{\hat{\sigma}^{2}_{D^{*}_{\Psi_{s}}}}{n} + (\hat{\Phi}_{s}(P_{n}))^{2}$

\end{table}

Table \ref{2} shows the bias, variance, mean of the estimated variance, mean squared error (MSE), $95\%$ confidence interval coverage, and power to detect the causal ATE (using $\alpha=0.05$) across 1000 iterations of this simulation. The standard CV-TMLE analyzed using the RCT data alone had nominal coverage of 0.95 and power of 0.64. The RCT-only CV-TMLE had higher power than any of the unadjusted estimators, with the RCT unadjusted estimator having coverage of 0.96 and power of 0.24. \par
The test-then-pool approaches were able to increase power as high as 0.93 for the TMLE-based test-then-pool when unbiased external data were available. However, as bias in the external data increased, the coverage suffered, dropping as low as 0.79 for the TMLE-based test-then-pool with $S=2$ and 0.76 for the t-test based test-then-pool with $S=3$. Test-then-pool is thus a high-risk, high-reward approach to integrating observational and RCT data.\par
Because the U-comparability assumption is not true, the two methods that rely only on a bias estimate of the ATE of A on the NCO also exhibited decreased coverage. Bias in the NCO-based difference-in-differences approach increased as the bias in the available external data increased, leading to coverage of 0.84 for the most biased external data dataset (S=3). When we only considered the estimated ATE of A on NCO in the ES-CVTMLE ($s_{n}^{\star\star\star}$ (\textbf{nco only})), coverage dropped as low as 0.87, which was lower coverage than when we also included $\hat{\Psi}^{\#}$ as an estimate of bias in the selector (discussed below). \par

 With default specifications, \textit{RBesT} \citep{weberRBesTBayesianEvidence2021} maintained coverage 0.94-0.97. Yet because this method does not adjust for covariates, power remained similar to the unadjusted estimator, with higher power achieved by considering external data with intermediate bias (power 0.32) than by considering unbiased external data (power 0.29). Thus, while \textit{RBest} resulted in close to nominal coverage, this method had lower power than alternative estimators, including an adjusted CV-TMLE using only the RCT data. MSE was higher for the \textit{RBesT} estimator than for any of the ES-CVTMLE estimators across all tested magnitudes of bias.\par
Next, we examine the ES-CVTMLEs with the \textbf{b2v} and \textbf{+nco} selectors. With $S=3$, these ES-CVTMLEs with selector $s_{n}^{\star}$ (\textbf{b2v}) or $s_{n}^{\star\star}$ (\textbf{+nco}) were approximately equivalent to the RCT-only CV-TMLE. This makes sense because data with bias this large was rejected, in which case the ES-CVTMLE algorithm is equivalent to a traditional CV-TMLE from the RCT only. When unbiased external controls were available, coverage was 0.96 for $s_{n}^{\star}$ (\textbf{b2v}) and $s_{n}^{\star\star}$ (\textbf{+nco}), suggesting somewhat conservative confidence intervals consistent with the fact that estimated bias is included in the limit distribution sampling procedure despite truly being zero. Power increased compared to the RCT-only CV-TMLE in either case but was lower with $s_{n}^{\star}$ (\textbf{b2v}) at 0.74, compared to 0.83 for $s_{n}^{\star\star}$ (\textbf{+nco}), demonstrating the utility of including information from the estimated ATE of A on the NCO in the selector for incorporating truly unbiased external controls. With $S=2$ (intermediate bias), coverage was 0.95 for $s_{n}^{\star}$ (\textbf{b2v}) and 0.92 for $s_{n}^{\star\star}$ (\textbf{+nco}), demonstrating that the ES-CVTMLE was able to maintain coverage close to 0.95 even with this challenging amount of bias and an imperfect NCO. \par
In this simulation, the ES-CVTMLE MSE was lower with either of the \textbf{b2v} or \textbf{+nco} selectors compared to the RCT-only CV-TMLE when considering $S=1$ and lower or the same when considering $S=2$ or $S=3$. Of all the compared estimators, the ES-CVTMLE with the $s_{n}^{\star}$ (\textbf{b2v}) selector provided the largest power gains with unbiased external data while maintaining $95\%$ coverage across all tested magnitudes of bias. However, the ES-CVTMLE with the $s_{n}^{\star\star}$ (\textbf{+nco}) selector is the estimator that decreased MSE the most when unbiased external data were available without increasing MSE when considering external data with intermediate or large bias. If we were running this simulation to choose an estimator for a proposed trial in a context when excessive randomization to control is considered unethical but we still desire greater protection against biased conclusions than a purely observational analysis could achieve, we might choose the ES-CVTMLE with the $s_{n}^{\star\star}$ (\textbf{+nco}) selector because it was able to boost power substantially when appropriate external controls were available while keeping coverage close to nominal for the simulated magnitudes of external bias, even when we did not have a perfect NCO. \par

\section{Real Data Application} \label{sec9}
To test the ability of the ES-CVTMLE to distinguish biased from unbiased external control data in a real data analysis, we use de-identified data from the LEADER trial \citep{marsoLiraglutideCardiovascularOutcomes2016}, which evaluated the effect of liraglutide (a medication for Type 2 Diabetes), on a primary combined outcome of cardiovascular death, nonfatal myocardial infarction, or nonfatal stroke. Because this trial evaluated a relatively rare binary outcome, the sample size was large enough to estimate the effect of liraglutide versus placebo on glycemic control (measured by hemoglobin A1c ($HbA_{1c}$)) with great precision. \par
\begin{figure}[H]
    \centering
    \includegraphics[width=\linewidth]{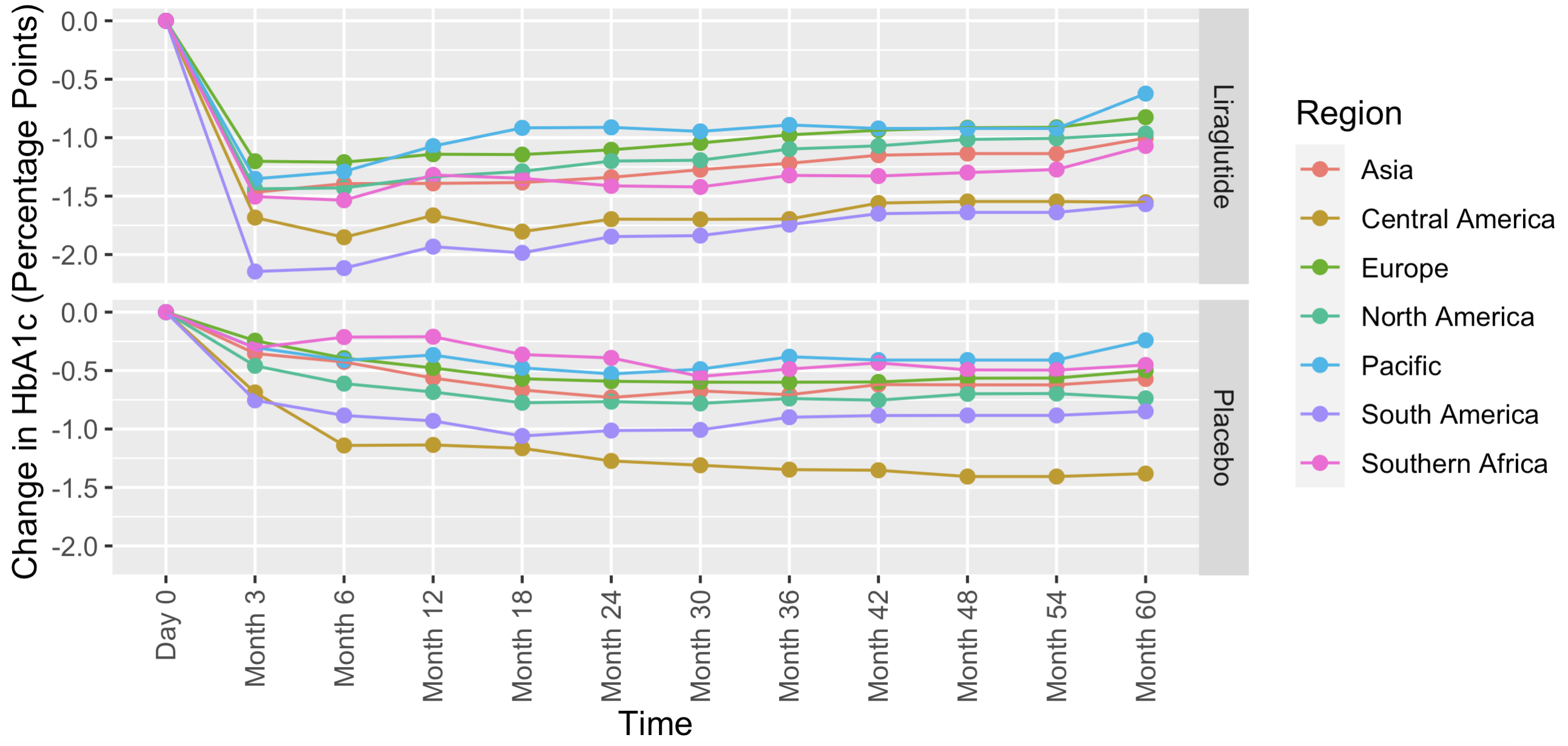}
    \caption{Change in $HbA_{1c}$ by Trial Arm and Region Over Time}
    \label{fig: 5}
\end{figure}

LEADER encouraged trial clinicians to optimize standard of care diabetes regimens beyond the addition of liraglutide or placebo to achieve a target $HbA_{1c}$ of $\leq 7\%$ for all participants \citep{zinmanLiraglutideGlycaemicOutcomes2018}. We thus would expect to see a change in $HbA_{1c}$ from baseline in the placebo arm due to modifications in baseline diabetes regimens, with larger improvements for participants with higher baseline $HbA_{1c}$. As shown in Figure \ref{fig: 5}, change in $HbA_{1c}$ differed by study region, with the largest average changes taking place in the Central and South American groups. Average baseline $HbA_{1c}$ was also higher in Central/South America ($9.29\%$) compared to Europe ($8.31\%$). \par
We can mimic a small RCT by taking a random sample of 150 participants from Central and South America (dataset ``CS'' below). To demonstrate the case where we would like to increase the number of patients receiving the intervention of interest in our ``RCT'', we select ``CS'' participants with a probability of 0.67 of having been in the liraglutide arm and 0.33 of having been in the placebo arm. If we were to augment the control arm of ``CS'' with 500 external controls from Central and South America (leading to hybrid dataset ``CS+''), we would expect those individuals who were randomized to placebo from within the same region to be unbiased controls. However, if we were to augment the small Central/South America RCT with 500 external controls from Europe (leading to hybrid dataset ``Eu+''), and if we treated baseline $HbA_{1c}$ as an unmeasured factor that causes the differences in placebo group outcomes by region, we would expect an overestimate of the effect of liraglutide on glycemic control compared to the effect estimate from the full Central/South America LEADER subset.  \par

\begin{figure}[H]
    \centering
\includegraphics[height=50mm]{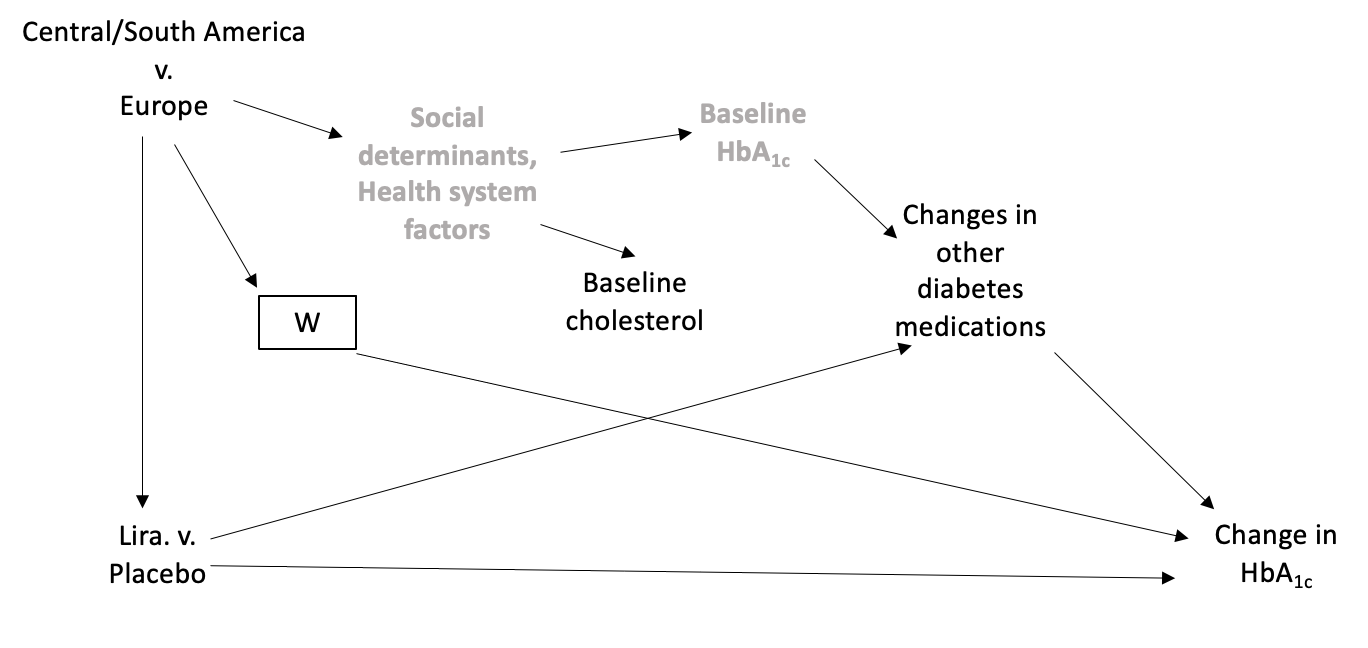}
    \caption{Causal Graph for Analysis of LEADER Data}
    \label{fig: 6}
    \footnotesize{\textbf{Caption:} Lira.: Liraglutide.}
\end{figure}
 This set-up implies the directed acyclic graph in Figure \ref{fig: 6}. Region affects treatment because members of the Central/South America group may receive liraglutide or placebo, and participants from Europe may only receive placebo. As noted above, region also affects change in $HbA_{1c}$ in the placebo arm. Higher average baseline $HbA_{1c}$ suggests that baseline diabetes regimens may have been, on average, less adequate in the Central/South America compared to European LEADER sample. In reviews of barriers and facilitators for diabetes management in Latin America, \citep{blasco-blascoBarriersFacilitatorsSuccessful2020} and \citep{aviles-santaCurrentStateDiabetes2020} cite access to healthcare, limitations in health system resources, and social determinants of health as challenges that impede optimal glycemic control for many people. While these factors vary by country, differences in underlying barriers between the Central/South American and European subgroups could explain at least part of the noted difference in average baseline $HbA_{1c}$. Baseline covariates ($W$) of age, sex, smoking status (never, former, or current), diabetes duration, whether the patient is insulin naive at baseline, eGFR, and BMI may also both differ by region and affect change in $HbA_{1c}$. Based on this DAG, we would expect baseline $HbA_{1c}$ and $W$ to block all paths from region to the outcome, other than the path through treatment, (satisfying \textbf{Assumption 2}), but we will treat baseline $HbA_{1c}$ as unmeasured. \par
Finally, we need an appropriate negative control. As shown in our causal graph, regional differences in health care causing inadequate control of $HbA_{1c}$ may also lead to inadequate control of cholesterol. This hypothesis is supported by Venkitachalam et al.'s \citep{venkitachalamGlobalVariationPrevalence2012} finding that both country-level health systems factors and economic development metrics were significantly associated with prevalence of elevated cholesterol among patients with a history of hyperlipidemia from thirty-six countries. If this hypothesis is true, baseline cholesterol may serve as a negative control variable given that it would be associated with unmeasured factors hypothesized to cause differences in the placebo arm change in $HbA_{1c}$ by region while not being affected by liraglutide administered post-baseline. Note that we expect improvements in the adequacy of the baseline medication regimen to lead to smaller improvements in $HbA_{1c}$ during the trial and also to be associated with lower levels of baseline cholesterol. By defining our outcome as improvement in $HbA_{1c}$, we satisfy our goal of defining a negative control variable that should be affected by the unmeasured bias in the same direction as the true outcome. \par
Our causal target parameter is the average treatment effect of liraglutide versus placebo on improvement in $HbA_{1c}$ from baseline to 12 months. Due to randomization within the LEADER trial, the ATE for the Central/South American context should be identifiable from CS and CS+ but not from Eu+ without adjustment for baseline $HbA_{1c}$. We compare the following estimators: a CV-TMLE for the full LEADER Central/South America dataset, a CV-TMLE using CS only, ES-CVTMLE considering CS+ or Eu+, and standard CV-TMLEs based on the CS+ and Eu+ datasets with no check for bias. We run this analysis 200 times with different random seeds for sampling from the full LEADER dataset. Further details regarding this analysis may be found in Appendix \ref{a9}.\par

\subsection{Results of Analysis of LEADER Data} \label{subsec9.1}
\begin{figure}
    \centering
 \includegraphics[width=\linewidth]{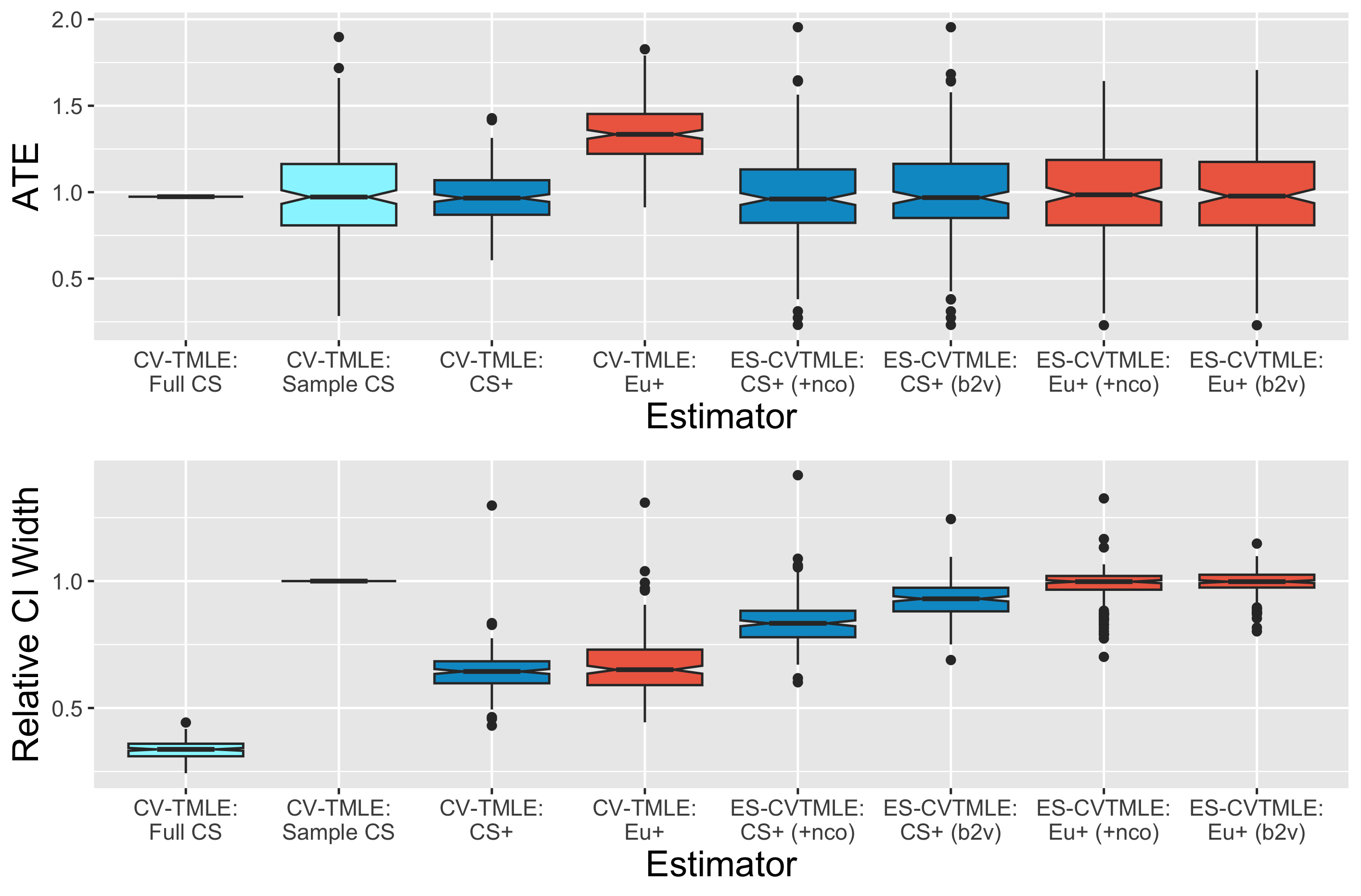}
    \caption{Estimated ATE of Liraglutide v. Placebo on Improvement in $HbA_{1c}$ by Estimator}
    \label{fig: 7}
    \footnotesize{\textbf{Caption:} Boxplots of ATE (average treatment effect) and relative confidence interval (CI) width. Relative width of CI compared to RCT CV-TMLE from sample CS. Full CS (light blue): Full Central/South America sample from LEADER trial (sample size 1182). CS (light blue): Central/South American sample ``RCT'' (sample size 150). CS+ (dark blue): CS plus 500 additional controls from Central/South America. Eu+ (orange): CS + 500 additional controls from Europe. CV-TMLE: Standard CV-TMLE from \textit{tmle} package \citep{gruberTmlePackageTargeted2012}}.
\end{figure}
Figure \ref{fig: 7} shows boxplots of the ATE point estimates and the relative confidence interval (CI) widths compared to the standard CV-TMLE from the small CS “RCT” sample for each estimator. The point estimate for the CV-TMLE from the full LEADER Central/South America subgroup was 0.97 percentage points. Due to the large sample size and 50-50 randomization, the width of the confidence interval from the full Central/South America CV-TMLE was a median of 0.34 relative to the confidence interval widths of the small CS sample CV-TMLE estimates. \par
The relative confidence interval widths for the CV-TMLEs using CS+ or Eu+ were intermediate, with medians of 0.64 and 0.65, respectively. Using a standard CV-TMLE for the CS+ datasets, the median point estimate was the same as for the full Central/South America LEADER trial at 0.97 percentage points. Yet without adjustment for baseline $HbA_{1c}$, the median ATE estimate for the standard CV-TMLE with Eu+ was severely biased at 1.33 percentage points. This example demonstrates what could happen in this analysis if we did not know about the regional differences in baseline HbA1c and decided to augment a Central/South America RCT with European controls with no analysis of bias. \par
Median (first and third quartile) values of the ATE estimates were 0.97 percentage points (0.81,1.16) for the CS-only CV-TMLE. ES-CVTMLE considering CS+ produced similar but slightly less variable results with median (first and third quartile) values of 0.97 percentage points (0.85,1.16) for the b2v selector and 0.96 percentage points (0.82,1.13) for the +nco selector. The median relative confidence interval width for the b2v selector was 0.93, while the median relative confidence interval width for the +nco selector was 0.83. As shown in the simulations, including the estimated ATE of treatment on a negative control outcome in the experiment selector can help to improve the efficiency of the estimator when unbiased external controls are available, yet the maximum improvement in confidence interval width we would expect is about half way between the CI width from the small RCT and the CI width of the pooled estimator; this is the price we pay for less biased estimates when only biased RWD is considered.  \par
With dataset Eu+, the median (first and third quartile) of the ATE estimates from the ES-CVTMLE were 0.98 percentage points (0.81,1.18) with the b2v selector and 0.98 percentage points (0.81,1.19) with the +nco selector.  ES-CVTMLE considering Eu+ had relative confidence interval widths that were similar to the CS-only standard CV-TMLE at a median of 1 for either selector. These results demonstrate the relative robustness of the ATE estimates from the ES-CVTMLE when potentially biased external controls are considered. 

\section{Discussion} \label{sec9}
We introduce a novel cross-validated targeted maximum likelihood estimator that aims to select the experiment (RCT or RCT plus external controls) with the best bias-variance tradeoff for the causal average treatment effect. To address the challenge that the selector may remain random asymptotically, we construct confidence intervals by sampling from the estimated limit distribution and by incorporating an estimate of the bias in this process. In a simulation mimicking a hybrid RCT-external data study, the ES-CVTMLE improved power compared to a standard CV-TMLE from the RCT only when unbiased external controls were available and maintained coverage close to $95\%$ with intermediate to large magnitudes of bias. In an analysis of the ATE of liraglutide versus placebo on improvement in 12 month $HbA_{1c}$ from the LEADER trial, we also demonstrated the ability of ES-CVTMLE to include external controls and narrow confidence intervals when additional unbiased controls are available and to reject biased external controls in the majority of iterations, maintaining similar confidence interval widths and point estimates compared to the sample ``RCT''-only CV-TMLE. \par
The purpose of the ES-CVTMLE is to provide an estimator that is relatively robust to varying magnitudes of bias in a combined RCT-external data analysis when we are not certain whether the necessary identification assumptions are true. Other data fusion estimators either rely on a comparison of mean outcomes or effect estimates for RCT versus external participants (e.g., \citep{vieleUseHistoricalControl2014, chengAdaptiveCombinationRandomized2021, oberstBiasrobustIntegrationObservational2022}) or evaluate the effect of treatment on an NCO (e.g., \citep{shiSelectiveReviewNegative2020}). We consider how both of these approaches may be useful together. Because bias must be estimated, attempts to optimize the bias-variance tradeoff may either exclude unbiased external data or include external data with a magnitude of bias that may impact causal coverage. ES-CVTMLE is able to incorporate information from both an estimated causal gap and from an NCO in the selector in an attempt to better distinguish biased from unbiased external controls. Yet we do not require the NCO to be perfect and show improved coverage compared to an NCO-based bias-adjustment approach when the U-comparability assumption is not true. Another advantage of the ES-CVTMLE is that it attempts to learn how much external information to include only from the data, rather than requiring a researcher to specify a level of confidence in the external controls as is required in some Bayesian dynamic borrowing approaches (e.g., \citep{schmidliRobustMetaanalyticpredictivePriors2014}) or to specify the value of a tuning parameter as is required by some frequentist approaches (e.g., \citep{yangElasticIntegrativeAnalysis2023, chenMinimaxRatesAdaptivity2021}). We note that adding the estimate of the ATE on the NCO to the bias term in the selector is one of many potential ways this information could be used to adjust the empirical bias-variance tradeoff. We explore the performance of the proposed selector as a proof of concept regarding how information from an NCO could be used to distinguish biased from unbiased external data in hybrid designs. Future work will consider alternate approaches to using an NCO in this context. \par
The largest limitation of the ES-CVTMLE is that we cannot guarantee $95\%$ CI coverage. This limitation is not unique to our estimator, as other data fusion estimators have demonstrated either increases in type 1 error or relative MSE or decreases in power with differing magnitudes of external data bias \citep{cuffeInclusionHistoricalControl2011, vieleUseHistoricalControl2014, galweySupplementationClinicalTrial2017, dejardinUseHistoricalControl2018,  harunCriticalAppraisalBayesian2020, yangElasticIntegrativeAnalysis2023, chenMinimaxRatesAdaptivity2021,   chengAdaptiveCombinationRandomized2021, oberstBiasrobustIntegrationObservational2022}. The first reason that coverage could be less than nominal is that, despite the goal of only including external data if doing so would improve the mean squared error, it is possible that external data integration would lead to decreased coverage for the causal target parameter. The second reason is that the statistical estimand is a data-adaptive target parameter, the limit distribution of which depends on the bias. We address the second challenge in two ways. First, as suggested by Leeb and Potscher \citep{leeb_model_2005}, our procedure for constructing confidence intervals treats the bias as a parameter that depends on sample size and therefore cannot be considered negligible in the limit distribution. Second, we use cross-validation to separate the data used to define the target parameter from the data used to estimate it. Cross-validated TMLE has shown good coverage for other DATPs in contexts such as estimation of the effect of an optimal dynamic treatment regime \citep{van_der_laan_targeted_2015, montoya_estimators_2023}. Nonetheless, the performance of the ES-CVTMLE and of other data fusion estimators may depend on characteristics of the proposed analysis, such as different RCT and external data sample sizes, the relative predictiveness of the covariates for the outcome, and the outcome type. For this reason, it would be important to conduct an outcome-blind simulation that is as true to a proposed study as possible, prior to implementing this estimator in a different context. It is possible that, for a given study design, the optimal bias-variance tradeoff across varying magnitudes of potential bias could be achieved by adding a smaller number of external controls than are available. A future version of the selector could consider adding different numbers of external controls or defining weighted combinations of RCT and combined experiment effect estimates. In future work, we also intend to evaluate the ES-CVTMLE in a wider variety of contexts, including extending the methods to include time-to-event outcomes. \par
We chose to implement a cross-validated estimator because CV-TMLE has outperformed non-cross-validated estimators in past work on data-adaptive target parameter estimation \citep{van_der_laan_targeted_2015, montoya_estimators_2023}. However, alternate approaches exist. For example, Chen et al. demonstrate the use of ensemble bagged estimators and the technique of sub-sampling without replacement to avoid the Donsker class condition that often motivates the use of cross-validation \citep{chen_debiased_2022}. Future work could consider extending such an approach for the context of data-adaptive target parameter estimation in general and for the use case of hybrid RCT-external data studies in particular. Another important area for future work is to develop guidance to help practitioners choose the optimal number of cross-validation folds for ES-CVTMLE in a given context. \par
 When considering future applications of the ES-CVTMLE methodology, we would expect bias to be smallest (and external data integration to be most likely) if the hybrid RCT-external data study is prospectively specified such that protocols and measurements are as similar as possible across data sources. We do not intend for these methods to replace a traditional RCT when it is feasible to run one for the sake of evaluating the efficacy of a new drug that has yet to be approved. Yet we hope that the ES-CVTMLE may ultimately provide evidence to support conclusions from underpowered RCTs conducted for rare diseases, allow randomization of more patients with severe diseases with few treatment options to an active treatment arm, and contribute evidence to the evaluation of previously approved drugs for new populations and indications.\\

\appendix

\setcounter{table}{0}
\renewcommand{\thetable}{S\arabic{table}}

\section{Table of Symbols} \label{a1}
\begin{table}[H]
\caption{Table of Symbols}
\label{s1}
\begin{tabular}{p{0.1\linewidth} | p{0.85\linewidth}}
\textbf{Symbol} & \textbf{Meaning}\\
\midrule
S & Variable indicating study in which an individual participated\\
X & Experiment. I.e., RCT alone is $X=0$, and RCT combined with external dataset $s$ is $X=s$.\\
A & Treatment\\
W & Baseline covariates\\
Y & Outcome\\
U & Exogenous variables\\
$O^n$ & Observed data $O^n=(O_1,...,O_n)$\\
\midrule
$P_{U,O}$ & True distribution of full data (endogenous and exogenous variables)\\
$P_0$ & True distribution of observed data\\
$P_n$ & Empirical distribution of observed data\\
$P_{n,v}$ & Empirical distribution of estimation set for cross-validation fold $v$\\
$P_{n,v^c}$ & Empirical distribution of experiment-selection set for cross-validation fold $v$\\
$P_{0,n}$ & True data distribution dependent on $n$\\
 \midrule
 $\Psi^{F}_{s}(P_{U,O})$ & $E_{W|X=s}[E(Y^1-Y^0|W,X=0)]$ (causal parameter: ATE)\\
 $\Psi_{s}(P_0)$ & $E_{W|X=s}[E_0[Y|A=1, W, X=0] - E_0[Y|A=0, W, X=s]]$ (statistical estimand for ATE)\\
 $\Phi_{s}(P_0)$ & $E_{W|X=s}[E_{0}[NCO|A=1, W, X=s] - E_{0}[NCO|A=0, W, X=s]]$ (statistical estimand for ATE on NCO)\\
 $\Psi^{\#,F}_{s}(P_{U,O})$ & $\Psi_{s}(P_0) - \Psi^{F}_{s}(P_{U,O})$ (the ``causal gap'' \citep{gruberTargetedLearningFuture2023})\\
 $\Psi^{\#}_{s}(P_0)$ & $E_{W|X=s}[E_{0}[Y|A=0,W,X=0]] - E_{W|X=s}[E_{0}[Y|A=0,W,X=s]]$ (statistical estimand for the causal gap)\\
 
\midrule
$s_{n}^{\star}$ & $\underset{s}{argmin}\frac{\hat{\sigma}^{2}_{D^{*}_{\Psi_{s}}}}{n} + (\hat{\Psi}^{\#}_{s}(P_{n}))^{2}$  (Bias$^2$ + variance selector \textbf{``b2v''}) \\
$s_{n}^{\star\star}$ & $\underset{s}{argmin}\frac{\hat{\sigma}^{2}_{D^{*}_{\Psi_{s}}}}{n} + (\hat{\Psi}^{\#}_{s}(P_{n}) + \hat{\Phi}_{s}(P_{n}))^{2}$ (Selector including ATE on NCO \textbf{``+nco''})  \\
$s_{n}^{\star\star\star}$ & $\underset{s}{argmin}\frac{\hat{\sigma}^{2}_{D^{*}_{\Psi_{s}}}}{n} + (\hat{\Phi}_{s}(P_{n}))^{2}$ (Bias only estimated as ATE on NCO \textbf{``nco only''}) \\

\end{tabular}

\end{table}

\section{Identification Proof} \label{a2}
The target parameter for the experiment with $X=s$ is
\begin{center}
    $\Psi_{s}^{F}(P_{U,O}) = E_{W|X=s}[E(Y^{1}|W,X=0)-E(Y^{0}|W,X=0)]$
\end{center}

where for each $a \in \{0,1\}$, $Y^{a}$ is the counterfactual outcome \citep{neymanApplicationsTheorieProbabilites1923} an individual would have had if they had received treatment $A=a$. 
The assumptions needed for identifiability are:\\
\textbf{Assumption 1} (randomization in the trial): $E[Y^{a}|W,X=0,A=a] = E[Y^{a}|W,X=0]$ for each $a \in \{0,1\}$ (true by design).\\
\textbf{Assumption 2}
(equivalence of covariate-adjusted mean outcomes for trial and combined controls):\\
$$E_{W|X=s}[E(Y|W,A=0,X=0)] = E_{W|X=s}[E(Y|W,A=0,X=s)].$$
\textbf{Assumption 3} (positivity): \textit{$P(A=a|W=w, X=s) > 0$ for all $a \in \{0,1\}$ and all $w$ for which $P(W=w, X=s) > 0$.} This assumption is true in the RCT by design and may be satisfied for other experiments by removing external data controls whose $W$ covariates do not have support in the trial population. \par

Proof:\\
\begin{center}
    $\Psi_{s}^{F}(P_{U,O}) = E_{W|X=s}[E(Y^{1}|W, X=0)-E(Y^{0}|W,X=0)]$
\end{center}
\begin{equation} \label{eq1}
 = E_{W|X=s}[E(Y^{1}|A=1,W,X=0)-E(Y^{0}|A=0,W,X=0)]
\end{equation}
\begin{equation} \label{eq2}
 = E_{W|X=s}[E(Y|A=1,W,X=0)-E(Y|A=0,W,X=0)]
\end{equation}
\begin{equation} \label{eq3}
 = E_{W|X=s}[E(Y|A=1,W,X=0)-E(Y|A=0,W,X=s)]
\end{equation}
\begin{center}
    $=\Psi_s(P_0)$.
\end{center}

where (5) holds by \textbf{Assumption 1}, (6) holds by the definition of counterfactuals \citep{pearlConsistencyRuleCausal2010}, (7) holds by  \textbf{Assumption 2}, and by \textbf{Assumption 3}, $\Psi_s(P_0)$ is well-defined.

\section{Identification of the Causal Gap} \label{a3}

As described in Section \ref{sec3}, for the experiment with $X=s$, the causal gap, $\Psi^{\#,F}_{s}(P_{U,O})$, is identified from the observed data.
Proof:
\begin{center}
    $ \Psi^{\#,F}_{s}(P_{U,O}) = \Psi_{s}(P_0) - \Psi^{F}_{s}(P_{U,O}) $\\
    $ = E_{W|X=s}[E(Y|A=1,W,X=0)-E(Y|A=0,W,X=s)] - $\\
    $E_{W|X=s}[E(Y^{1}|W,X=0)-E(Y^{0}|W,X=0)]$
\end{center}
\begin{equation} \label{eq2}
\begin{split}
    = E_{W|X=s}[E(Y|A=1,W,X=0)-E(Y|A=0,W,X=s)] - \\
    E_{W|X=s}[E(Y^1|A=1,W,X=0)-E(Y^0|A=0,W,X=0)]
\end{split}
\end{equation}
\begin{equation} \label{eq3}
\begin{split}
    = E_{W|X=s}[E(Y|A=1,W,X=0)-E(Y|A=0,W,X=s)] - \\
    E_{W|X=s}[E(Y|A=1,W,X=0)-E(Y|A=0,W,X=0)]
\end{split}
\end{equation}
\begin{equation} \label{eq6}
\begin{split}
    = E_{W|X=s}[E(Y|A=0,
W,X=0)] - E_{W|X=s}[E(Y|A=0,W,X=s)]\\
\end{split}
\end{equation}
\begin{center}
    $=\Psi^{\#}_{s}(P_0)$
\end{center}

where (8) holds by \textbf{Assumption 1} due to randomization in the RCT, (9) holds by the definition of counterfactual outcomes \citep{pearlConsistencyRuleCausal2010}, (10) is a simplification of equation (9), and by \textbf{Assumption 3} of positivity applied to the RCT experiment (which is guaranteed to hold due to randomization in the trial) and the fact that all external controls receive $A=0$, $\Psi^{\#}_{s}(P_0)$ is well-defined.\par

\section{Assumptions Required for Equivalence of the Statistical Estimand for the ATE on an NCO and the Causal Gap} \label{a4}

Sofer et al. \citep{soferNegativeOutcomeControl2016}, Miao et al. \citep{miaoConfoundingBridgeApproach2020}, and Shi et al. \citep{shiSelectiveReviewNegative2020} discuss assumptions under which bias due to confounding is identified using a negative control outcome. Generalizing these assumptions slightly to capture the multiple potential sources of bias in our hybrid study, we note that the causal gap is identified by the statistical estimand for the ATE of treatment on the NCO under the following assumptions for a given experiment with $X=s$.\\

\textbf{Assumption 4}: $NCO \perp A | S, U_Y$. This assumption is an adaptation of the U-comparability assumption \citep{lipsitchNegativeControlsTool2010} for identifying bias due to unmeasured confounding using an NCO to this context where the potential sources of bias include unmeasured common causes of trial participation and outcomes and a direct effect of trial participation on outcomes, rather than traditional confounding bias. This assumption also implies that treatment does not directly affect the NCO, consistent with the definition of a negative control outcome. \par
\textbf{Assumption 5}: 
$$\Psi^{\#,F}_{s}(P_{U,O}) = E_{W|X=s}[E_0[Y|A=1, W, X=s] - E_0[Y|A=0, W, X=s]]$$ 
$$- E_{W|X=s}[E(Y^1-Y^0|W,X=s)]$$
$$ = E_{W|X=s}[E_{0}[NCO |A=1,W,X=s] - E_{0}[NCO |A=0, W, X=s]] = \Phi_{s}(P_0).$$ 
Similarly to the ``additive equi-confounding'' \citep{soferNegativeOutcomeControl2016} assumption for estimating bias due to confounding but adapted for the potential sources of bias inherent in this hybrid design, \textbf{Assumption 5} is more stringent than \textbf{Assumption 4} because it requires that the magnitude and direction of the total effect of all sources of bias on the treatment-NCO relationship is the same as on the treatment-primary outcome relationship. If this assumption is true, the statistical estimand for the ATE of treatment on the NCO, $\Phi_{s}(P_0)$, is equivalent to the causal gap for our target parameter, $\Psi^{\#,F}_{s}(P_{U,O})$. If \textbf{Assumption 4} is true but \textbf{Assumption 5} is false, then the causal gap may still be estimated based on an NCO, but more complex methods than simply estimating $\Phi_{s}(P_0)$ are required. For example, Miao et al. \citep{miaoConfoundingBridgeApproach2020} propose an approach that utilizes both an NCO and a negative control exposure. In this version of ES-CVTMLE, we estimate bias based on the NCO as $\Phi_{s}(P_0)$, but more complex methods of estimating bias using negative control variables may be considered in future versions of the estimator. \par

\section{Estimation of Bias} \label{a5}

In order to estimate the causal gap, $\Psi^{\#}_{s}(P_{0})$, we will use targeted maximum likelihood estimation \citep{vanderlaanTargetedMaximumLikelihood2006a}. The efficient influence curve (EIC) for
 \begin{center}
 $\Psi^{0}_{s}(P_{0}) = E_{W|X=s}[E_{0}[Y|A=0,W,X=s]]$ is\\
     $D^{*}_{\Psi^{0}_{s}}(O) = \frac{I(X=s)}{P_0(X=s)}( \frac{I(A=0)}{P_{0}(A=0|W,X=s)}(Y-E_{0}[Y|A,W,X=s])$\\
     $ + E_{0}[Y|A=0,W,X=s] - \Psi^{0}_{s}(P_{0})).$
 \end{center}
TMLE involves fitting initial estimates of the treatment mechanism, $\hat{P}_{n}(A=0|W,X=s)$, and outcome regression, $\hat{E}_{n}[Y|A,W,X=s]$, with the SuperLearner ensemble machine learning algorithm \citep{vanderlaanSuperLearner2007}. The initial estimate is then targeted using a parametric working model \citep{vanderlaanTargetedLearning2011}
\begin{center}
logit$(\hat{E}^{*}_{n}[Y|A,W,X=s] = $logit$(\hat{E}_{n}[Y|A,W,X=s]) + \epsilon_n H^{*}_{s,n}(A,W,X=s)$
\end{center}
where $H^{*}_{s,n}(A,W,X=s) = \frac{I(A=0,X=s)}{\hat{P}_{n}(A=0|W,X=s)P_{n}(X=s)}$ is the covariate in front of the residual in the EIC, and $\epsilon_n$ may be fitted using logistic regression of $Y$ on $H^{*}_{s,n}$ with offset logit$(\hat{E}_{n}[Y|A,W,X=s])$ among observations with $X=s$. While a linear regression may be performed for a continuous outcome, it is common practice to scale the outcome as $(Y - min(Y))/(max(Y) - min(Y)))$, perform the TMLE with a logistic fluctuation, and re-scale the parameter estimate to the original scale in order to respect the bounds of the observed data distribution \citep{gruberTargetedMaximumLikelihood2010a}. \par

The conditional expectation of the outcome under control is updated as
\begin{center}
logit$(\hat{E}^{*}_{n}[Y|A=0,W,X=s]) = $logit$(\hat{E}_{n}[Y|A=0,W,X=s]) + \epsilon_n H^{*}_{s,n}(0,W,X=s)$
\end{center}
where $H^{*}_{s,n}(0,W,X=s) = \frac{1}{\hat{P}_{n}(A=0|W,X=s)P_{n}(X=s)}$ using the same $\epsilon_n$. The final estimate of the mean outcome under control in the combined dataset is then
\begin{center}
$\hat{\Psi}^{0}_{s}(P_{n}) = \frac{1}{n}\sum_{i=1}^{n}\frac{I(X_i =s)}{P(X=s)}\hat{E}^{*}_{n}[Y|A_i=0,W_i,X_i = s]$.
\end{center}

An alternate option that may be more stable in the context of near-violations of the positivity assumption is to move the denominator of the clever covariate to the denominator of the weights for the regression training the TMLE coefficient \citep{robinsCommentPerformanceDoubleRobust2007, rotnitzkyImprovedDoublerobustEstimation2012, tranDoubleRobustEfficient2019}. For this option of ``targeting the weights'', we perform a logistic regression of binary or scaled-continuous $Y$ on $H^{*}_{s,n}(A,W,X=s)=I(A=0, X=s)$ with offset $logit(\hat{E}_{n}[Y|A,W,X=s])$ and weights $\frac{I(A=0,X=s)}{\hat{P}_{n}(A=0|W,X=s)P_{n}(X=s)}$ among observations with $X=s$. Initial estimates are then updated as $\hat{E}^{*}_{n}[Y|A=0,W,X=s] = logit^{-1}(logit(\hat{E}_{n}[Y|A=0,W,X=s]) + \epsilon_{n})$.

We can also use TMLE to estimate $\tilde{\Psi}^{0}_{s}(P_{0}) = E_{W|X=s}[E_{0}[Y|A=0,W,X=0]]$, with EIC

\begin{center}
    $D^{*}_{\tilde{\Psi}^{0}_{s}}(O) = \frac{I(X=s)}{P_0(X=s)}(\frac{I(S=0,A=0)}{P_{0}(S=0|A=0,W,X=s)P_{0}(A=0|W,X=s)}
(Y - E_{0}[Y|A,W,S,X=s])$\\
$+ E_{0}[Y|A=0,W,S=0,X=s] - \tilde{\Psi}^{0}_{s}(P_{0})).$
\end{center}

We use the same procedure as above with clever covariate\\
$H^{*}_{s,n}(A,W,S,X=s) = \frac{I(S=0,A=0)}{\hat{P}_{n}(S=0|A=0,W,X=s)\hat{P}_{n}(A=0|W,X=s)P_{n}(X=s)}$ to obtain a targeted estimate $\hat{E}^{*}_{n}[Y|A=0,W,S=0,X=s]$. Our updated estimate\\ $\hat{\tilde{\Psi}}^{0}_{s}(P_{n}) = \frac{1}{n}\sum_{i=1}^{n}\frac{I(X=s)}{P_{n}(X=s)}\hat{E}^{*}_{n}[Y|A_i=0,W_i,S_i=0,X_i = s]$. Then, our TMLE estimate of the causal gap is
$$\hat{\Psi}^{\#,TMLE}_{s}(P_{n}) = \hat{\tilde{\Psi}}^{0}_{s}(P_{n}) - \hat{\Psi}^{0}_{s}(P_{n}).$$
The statistical estimand for the ATE on the NCO may be estimated using a standard TMLE for the average treatment effect \citep{vanderlaanTargetedLearning2011}, yielding $\hat{\Phi}_{s}^{TMLE}(P_n)$.

\section{Algorithm for ES-CVTMLE} \label{a6}
\begin{figure}[H]
    \centering
    \includegraphics[height=200mm]{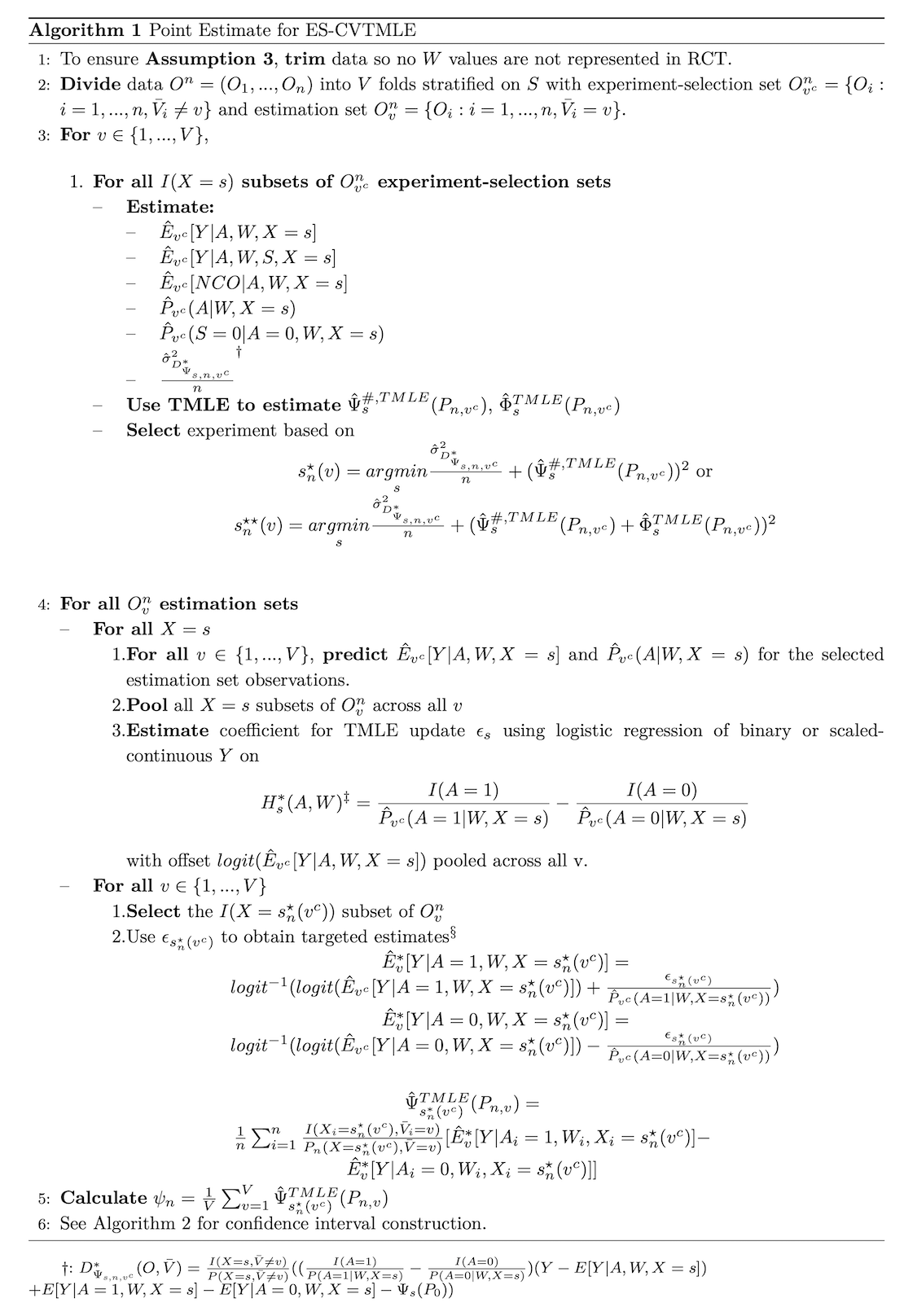}
    \label{alg: 1}
     \footnotesize{$\ddagger$: An alternative method that may be more stable in the context of practical positivity violations is to ``target the weights'' 
\citep{robinsCommentPerformanceDoubleRobust2007, rotnitzkyImprovedDoublerobustEstimation2012, tranDoubleRobustEfficient2019} by using clever covariate $H^{*}_{s}(A,W) = I(A=1) - I(A=0)$ and weights $\frac{I(A=1)}{\hat{P}_{v^c}(A=1|W,X=s)} + \frac{I(A=0)}{\hat{P}_{v^c}(A=0|W,X=s)}$. \\
$\S$: Re-scale $\hat{E}^{*}_{v}[Y|A, W, X=s^{\star}_{n}(v^c)]$ to the original outcome scale if using scaled-continuous Y.}
\end{figure}

\normalsize

\begin{figure}[H]
    \centering
    \includegraphics[height=200mm]{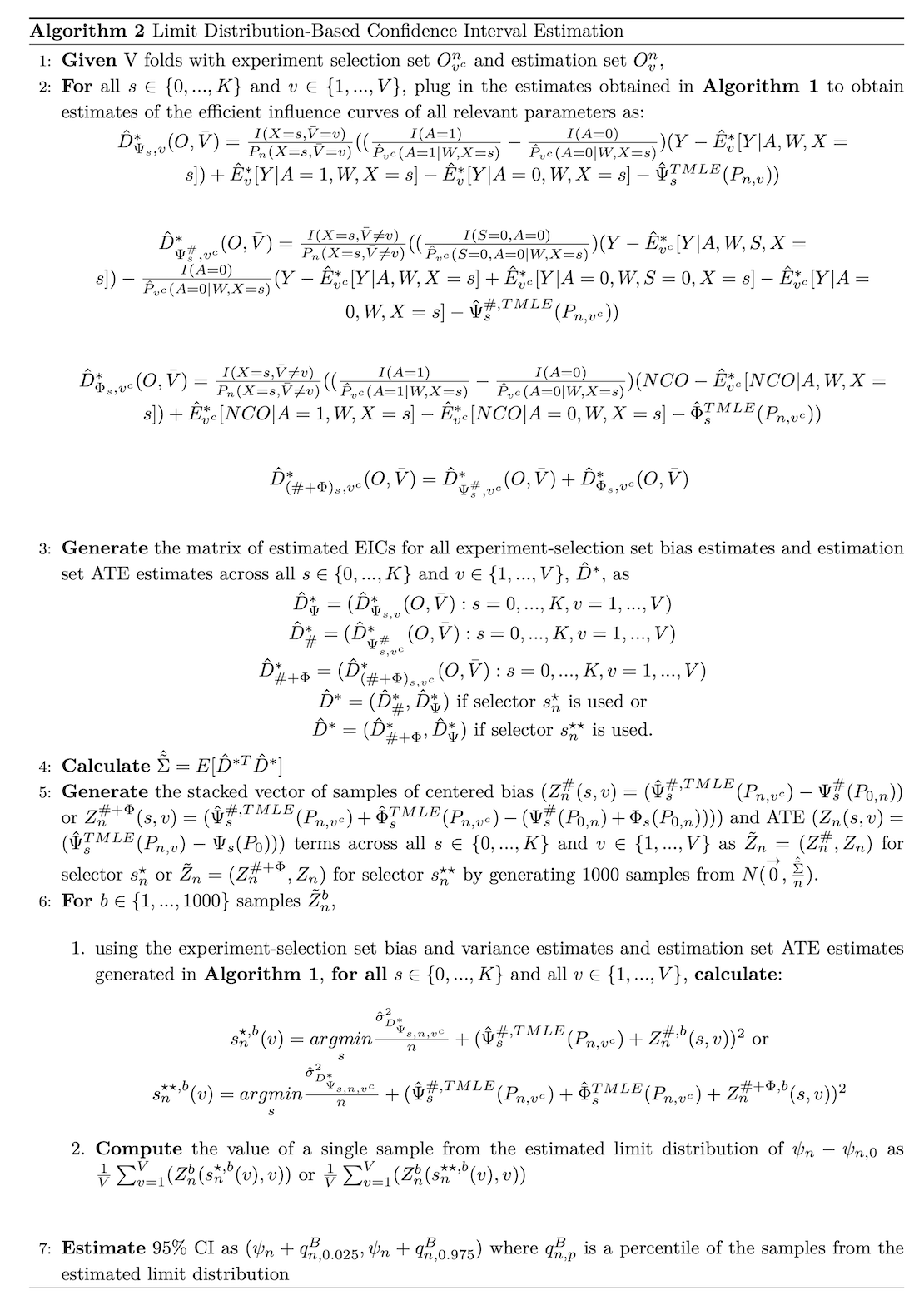}
    \label{alg2: 1}
\end{figure}

Note that $\hat{E}^{*}_{v^c}$ denotes training and TMLE targeting of the conditional expectation on experiment-selection set data, as is done for estimation of the bias terms.
If RCT-only is selected in all cross-validation folds, the ES-CVTMLE is equivalent to a standard CV-TMLE from the RCT only, with confidence intervals estimated as \citep{zhengAsymptoticTheoryCrossValidated2010, hubbardStatisticalInferenceData2016}: $\psi_{n} \pm 1.96*(\frac{1}{V}\sum_{v=1}^{V}\frac{\hat{\sigma}^{2}_{D^{*}{X=0,n,v}}}{n_{X=0}})^{1/2}$ where 
\begin{center}
    $D^{*}_{X=0,n,v} = (\frac{I(A=1)}{P(A=1|W,X=0)} - \frac{I(A=0)}{P(A=0|W,X=0)})(Y-E[Y|A,W,X=0])$\\
    $+ E[Y|A=1,W,X=0] - E[Y|A=0,W,X=0] - \Psi_{X=0}(P_0)$
\end{center}

estimated among RCT estimation set observations for fold $v$. 

\section{Proof of Theorem 1} \label{a7}
Let $P_{n,v}^{*}$ denote training of initial estimators of the outcome regression $\hat{E}_{v^c}[Y|A,W,X=s]$ and treatment mechanism $\hat{P}_{v^c}(A|W)$ on experiment selection sets with TMLE targeting on separate or pooled estimation sets to generate $\hat{E}^{*}_{v}[Y|A,W,X=s]$ for fold $v$. Let $Q^{*}_{n,v}(A,W,X=s) = \hat{E}^{*}_{v}[Y|A,W,X=s]$ and $Q_{0}(A,W,X=s) = E_{0}[Y|A,W,X=s]$. Finally, let $P_{n,v^c}^{*}$ denote training and TMLE targeting of the relevant outcome regressions for each bias parameter on experiment selection sets for fold $v$.

\textbf{Conditions for Theorem 1:}
\begin{enumerate}
    \item \textbf{Convergence of Second-Order Remainders}
    \begin{enumerate}
        \item Second-order remainder for ES-CVTMLE: $\frac{1}{V}\sum_{v=1}^{V}(R_{s}(P_{n,v}^{*},P_0)) = o_P((n_{X=s})^{-1/2})$.\\
        $R_{s}(P_{n,v}^{*},P_0) = R_{s,1}(P_{n,v}^{*},P_0) - R_{s,0}(P_{n,v}^{*},P_0)$ where for each $s \in 0,...,K$
   \begin{center}
    $R_{s,a}(P_{n,v}^{*},P_0)= E_P\{\frac{1}{\hat{P}_{v^c}(a|X=s,W)}(Q^{*}_{n,v} - Q_{0})(a,W,X=s)$\\
    $(\hat{P}_{n,v^c}- P_{0})(a|W,X=s) \} $\\
    \end{center}
    \item Second-order remainder for $Z^{\#}_{n}(s,v)$: $R_{\#,s}(P_{n,v^c}^{*},P_0) = o_P((n_{X=s})^{-1/2})$ for each $s \in 0,...,K$, $v \in 1,...,V$ where
    \begin{center}
        $R_{\#,s}(P_{n,v^c}^{*},P_0) = \Psi^{\#}_s(P_{n,v^c}^{*}) - \Psi^{\#}_{s}(P_0)+P_0D^{*}_{\Psi^{\#}_s}(P_{n,v^c}^{*})$
    \end{center}
    \item Second-order remainder for $Z^{\#+\Phi}_{n}(s,v)$:\\
    $R_{(\#+\Phi),s}(P_{n,v^c}^{*},P_0) = o_P((n_{X=s})^{-1/2})$ for each $s \in 0,...,K$, $v \in 1,...,V$ where
    \begin{center}
        $R_{\#+\Phi,s}(P_{n,v^c}^{*},P_0) = (\Psi^{\#}_s+ \Phi_s)(P_{n,v^c}^{*}) - (\Psi^{\#}_{s}+ \Phi_{s})(P_0)+P_0D^{*}_{(\#+\Phi)_{s}}(P_{n,v^c}^{*})$
    \end{center}
    \end{enumerate}
    \item \textbf{Consistency of EIC Estimation}: For each $s \in 0,...,K$, $v \in 1,...,V$:\\ $P_0\{ D^*_{\Psi_{s},v}(P_{n,v}^{*})-D^*_{\Psi_{s},v}(P_0)\}^2 \overset{p}{\rightarrow} 0$, $P_0\{ D^*_{\Psi^{\#}_{s},v^c}(P_{n,v^c}^{*})-D^*_{\Psi^{\#}_{s},v^c}(P_0)\}^2 \overset{p}{\rightarrow} 0$,\\
    and $P_0\{ D^*_{(\#+\Phi)_{s},v^c}(P_{n,v^c}^{*})-D^*_{(\#+\Phi)_{s},v^c}(P_0)\}^2\overset{p}{\rightarrow} 0$
    \item \textbf{Donsker Class Condition for Bias Term Estimation}: $\{D^{*}_{\Psi^{\#}_{s},v^c}(P):P \in M\}$ and $\{D^{*}_{(\#+\Phi)_{s},v^c}(P):P \in M\}$ are $P_0$-Donsker, where $M$ defines the set of possible distributions $P$ \citep{vanderlaanTargetedLearning2011}.
    
\end{enumerate}
\textbf{Proof of Theorem 1:}

\begin{center}
    $\sqrt{n}(\psi_{n}-\psi_{n,0})=\frac{\sqrt{n}}{V}\sum_{v=1}^{V} (\hat{\Psi}_{s^{\star}_{n}(v^c)}(P_{n,v}^{*})-\Psi_{s^{\star}_{n}(v^c)}(P_0))=
\frac{\sqrt{n}}{V}\sum_{v=1}^{V}((P_{n ,v}-P_0)D^{*}_{s^{\star}_{n}(v^c)}(P_{n,v}^{*})+ R_{s^{\star}_{n}(v^c)}(P_{n,v}^{*},P_0))$
$=
\frac{\sqrt{n}}{V}\sum_{v=1}^{V}((P_{n ,v}-P_0)D^{*}_{s^{\star}_{n}(v^c)}(P_{0}) + (P_{n ,v}-P_0)(D^{*}_{s^{\star}_{n}(v^c)}(P_{n,v}^{*}) - D^{*}_{s^{\star}_{n}(v^c)}(P_{0}))+ R_{s^{\star}_{n}(v^c)}(P_{n,v}^{*},P_0))$
$=
\frac{\sqrt{n}}{V}\sum_{v=1}^{V}((P_{n ,v}-P_0)D^{*}_{s^{\star}_{n}(v^c)}(P_{0})) + o_P(1)$
\end{center}
by assumption of \textit{Conditions 1 and 2}. Define
\begin{center}
    $Z^{\dagger}_{n}(s,v) = \sqrt{n}(P_{n ,v}-P_0)D^{*}_{s}(P_{0})$
\end{center}
 By the Central Limit Theorem, across all $s$ and $v$, the vector $Z^{\dagger}_{n} = (Z^{\dagger}_{n}(s,v):s,v) \overset{D}{\rightarrow} N(\overset{\rightarrow}{0}, \Sigma^{\Psi})$.

In order to understand the behavior of $\sqrt{n}(\psi_{n}-\psi_{n,0})$, we must also understand the behavior of $s_{n}^{\star}(v)$, which depends on the behavior of either $Z^{\#}_{n}(s,v)$ or $Z^{\# + \Phi}_{n}(s,v)$.

For the standardized bias terms estimated on experiment-selection sets,
\begin{center}
    $Z^{\#}_{n}(s,v) = \sqrt{n}(\hat{\Psi}^{\#}_{s}(P_{n,v^c}^{*}) - \Psi^{\#}_{s}(P_{0}))$\\
    $=\sqrt{n}((P_{n,v^c} - P_0)D^{*}_{\Psi^{\#}_{s},v^c}(P_{n,v^c}^{*}) + R_{\#,s}(P_{n,v^c}^{*},P_0))$\\
    $=\sqrt{n}((P_{n,v^c} - P_0)D^{*}_{\Psi^{\#}_{s},v^c}(P_0)
    + (P_{n,v^c} - P_0)\{D^{*}_{\Psi^{\#}_{s},v^c}(P_{n,v^c}^{*}) - D^{*}_{\Psi^{\#}_{s},v^c}(P_0)\}
    + R_{\#,s}(P_{n,v^c}^{*},P_0))$\\
    $=\sqrt{n}(P_{n,v^c} - P_0)D^{*}_{\Psi^{\#}_{s},v^c}(P_0) + o_P(1)$
\end{center}
by assumption of \textit{Conditions 1, 2, and 3}. If $\hat{\Phi}_{s}$ is include in the bias estimation, where $D^{*}_{(\#+\Phi)_{s},v^c} = D^{*}_{\Psi^{\#}_{s},v^c} + D^{*}_{\Phi_{s},v^c}$,

\begin{center}
    $Z^{\# + \Phi}_{n}(s,v) = \sqrt{n}((\hat{\Psi}^{\#}_{s} + \hat{\Phi}_{s})(P_{n,v^c}^{*}) - (\Psi^{\#}_{s}+\Phi_{s})(P_{0}))$\\
    $=\sqrt{n}((P_{n,v^c} - P_0)D^{*}_{(\#+\Phi)_{s},v^c}(P_{n,v^c}^{*}) + R_{(\#+\Phi),s}(P_{n,v^c}^{*},P_0))$\\
    $=\sqrt{n}((P_{n,v^c} - P_0)D^{*}_{(\#+\Phi)_{s},v^c}(P_0)
    + (P_{n,v^c} - P_0)\{D^{*}_{(\#+\Phi)_{s},v}(P_{n,v^c}^{*}) - D^{*}_{(\#+\Phi)_{s},v^c}(P_0)\}
    + R_{(\#+\Phi),s}(P_{n,v^c}^{*},P_0))$\\
    $=\sqrt{n}(P_{n,v^c} - P_0)D^{*}_{(\#+\Phi)_{s},v^c}(P_0) + o_P(1)$ 
\end{center}
by assumption of \textit{Conditions 1, 2, and 3}. \par
By the Central Limit Theorem, $Z_{n}^{\#}(s,v)$ and $Z_{n}^{\#+\Phi}(s,v)$ also converge to normal distributions. Across all $s$ and $v$, 
    \begin{center}
    $Z_{n}^{\#}=(Z_n^{\#}(s,v): s,v) \overset{D}{\rightarrow} Z^{\#} \sim N(\overset{\rightarrow}{0}, \Sigma^{\#})$\\ $Z_{n}^{\#+\Phi}=(Z_n^{\#+\Phi}(s,v): s,v) \overset{D}{\rightarrow} Z^{\#+\Phi} \sim N(\overset{\rightarrow}{0}, \Sigma^{\#+\Phi})$\\
        $\tilde{Z_n} = (Z^{\#}_n,Z^{\dagger}_n) \overset{D}{\rightarrow} \tilde{Z} \sim N(\overset{\rightarrow}{0},\tilde{\Sigma})$\\
        or $\tilde{Z}_n = (Z^{\#+\Phi}_n,Z^{\dagger}_n) \overset{D}{\rightarrow} \tilde{Z} \sim N(\overset{\rightarrow}{0},\tilde{\Sigma})$\\
        where $\tilde{\Sigma}$ is defined in Section \ref{subsec6.1}. The limit distribution of the experiment-selector CV-TMLE is then defined by sampling from $\tilde{Z}$, calculating 
        
        \begin{center}
    $\bar{s}^{\star}(v^c) = \underset{s}{argmin} \hspace{0.1cm} \sigma^{2}_{D^{*}_{\Psi_{s},v^c}}  + (Z^{\#}(s,v) + \sqrt{n}\Psi^{\#}_{s}(P_{0}))^{2} $\\
or \\
  $\bar{s}^{\star\star}(v^c)  = \underset{s}{argmin} \hspace{0.1cm} \sigma^{2}_{D^{*}_{\Psi_{s},v^c}}  + (Z^{\#+\Phi}(s,v) + \sqrt{n}(\Psi^{\#}_{s}(P_{0}) + \Phi_{s}(P_{0})))^{2} $\\
\end{center}
and finally calculating
        \begin{center}
            $\sqrt{n}(\psi_{n}-\psi_{n,0}) = \frac{\sqrt{n}}{V}\sum_{v=1}^{V}(P_{n,v}-P_0)D^{*}_{\Psi_{\bar{s}^{\star}(v^c)}}(P_0) + o_{P}(1) = \frac{1}{V}\sum_{v=1}^{V}(Z^{\dagger}(\bar{S}^{\star}(v^c),v)) + o_{P}(1)$\\
        \end{center}
         
    or 
    
     \begin{center}
            $\sqrt{n}(\psi_{n}-\psi_{n,0}) = \frac{\sqrt{n}}{V}\sum_{v=1}^{V}(P_{n,v}-P_0)D^{*}_{\Psi_{\bar{s}^{\star\star}(v^c)}}(P_0) + o_{P}(1) = \frac{1}{V}\sum_{v=1}^{V}(Z^{\dagger}(\bar{S}^{\star\star}(v^c),v)) + o_{P}(1)$.\\
        \end{center}
        
    \end{center}
    
    $\sqrt{n}(\psi_{n}-\psi_{n,0})$ thus converges to an average of mixtures of normal distributions.

\section{Data Generating Process for Simulation} \label{a8}

As described in Section \ref{subsec8.1}, four datasets were simulated as follows: 1) an ``RCT'' dataset of 150 observations with $S=0$, $A=1$ randomized with probability 0.67, and bias terms $B_1$ and $B_2$ equal to zero, 2) an external control dataset of 500 observations with $S=1$, $A=0$, and bias terms $B_1$ and $B_2$ equal to zero, 3) an external control dataset of 500 observations with $S=2$, $A=0$, and $B_1 + B_2 \approx B$, and 4) an external control dataset of 500 observations with $S=3$, $A=0$, and $B_1 + B_2 \approx 5*B$. For this simulation, biased external data could be included if the magnitude of bias is approximately  $\sqrt{\frac{\hat{\sigma}^{2}_{D^{*}_{\Psi_{X=0,n,v^c}}}}{n} - \frac{\hat{\sigma}^{2}_{D^{*}_{\Psi_{X=2,n,v^c}}}}{n}} = B = 0.21$. We generate $B_{1}$ and $B_{2}$ as described below:

\begin{table}[H]
\begin{center}
\caption{Bias Terms for Simulated Datasets}
\label{S2}
\begin{tabular}{@{}lll@{}}
Dataset & $B_1$ & $B_2$\\
\midrule
 $S=0$ & 0 & 0\\
 $S=1$ & 0 & 0\\
 $S=2$ & $N(\frac{3}{4}B, 0.02^2)$ & $N(\frac{1}{4}B, 0.02^2)$\\
 $S=3$ & $N(\frac{3}{4}*5*B, 0.02^2)$ & $N(\frac{1}{4}*5*B, 0.02^2)$\\
\end{tabular}
\end{center}
\end{table}

We also simulate two covariates, W1 and W2, as $N(0,1)$. The outcome $Y$ and $NCO$ are then simulated as
\begin{center}
    $Y = -3 + 2*W1 + W2 - 0.6*A + B_1 + B_2 + U_Y$\\
  $ NCO = -2 + W1 + 2*W2 + B_1 + U_{nco}$
\end{center}

  with $U_Y \sim N(0,1.5^2)$ and $U_{nco} \sim N(0,1.5^2)$. The true causal ATE of A on Y  in this simulation is $-0.6$. The NCO is affected by $B_1$ but not $B_2$, and so the U-comparability and additive equi-confounding assumptions do not completely hold in this case. \par
 
 For the TMLE-based methods used in the simulation, we use linear regression for the outcome regression and a candidate library for the treatment mechanism consisting of lasso regression \citep{friedmanRegularizationPathsGeneralized2010} or the mean. We use 10-fold cross-validation and target the weights as described above. When only $S=0$ data are considered, we use the true randomization probability for $P(A=1|W)$.

\section{Further Details Regarding Real Data Analysis} \label{a9}

In our real data analysis of the Central/South America and European subsets of the LEADER trial, overall missingness for change in $HbA_{1c}$ was $6\%$. Missingness for baseline cholesterol, which was treated as an outcome for the estimate of the ATE of A on negative control in the selector but was treated as a baseline variable in the TMLE for the ATE of A on Y, was $2\%$. Outcome missingness was handled with inverse probability weights, consistent with the \textit{tmle} package \citep{gruberTmlePackageTargeted2012}. Specifically, we define a binary variable $\Delta$ that indicates an outcome was not missing. Clever covariates for all TMLEs were then modified to include the missingness indicator in the numerator and missingness mechanism in the denominator. For example, the clever covariate for the ATE was modified as $H(A,W)=\frac{\Delta(2*A-1)}{P(\Delta=1|A,W)P(A|W)}$. Missingness for baseline covariates, which was less than $0.1\%$ for all $W$ variables, was imputed using the \textit{R} package \textit{mice: Multivariate Imputation by Chained Equations} \citep{vanbuurenMiceMultivariateImputation2011} separately for each study.\par

For the TMLEs, we use the following specifications. We employ a discrete Super Learner for all outcome regressions with a library consisting of linear regression \citep{eneaSpeedglmFittingLinear2021}, lasso regression (via \textit{R} package $\textit{glmnet}$ \citep{friedmanRegularizationPathsGeneralized2010}), and multivariate adaptive regression splines \citep{milborrowEarthMultivariateAdaptive2020}. When considering only $S=1$, we use the true randomization probability of 0.67 for $P(A=1)$. When external controls are considered, we use a discrete Super Learner with library consisting of logistic regression and lasso regression for the treatment mechanism. Because missingness was low, for the missingness mechanism we use a linear model adjusting only for treatment unless the number of missing observations is less than five, in which case we employ an intercept only adjustment. We also use the \textit{tmle} package defaults of fitting a CV-TMLE, using a logistic fluctuation, and targeting the weights, as described above.

\section{Author Statements}

\textbf{Funding Information}

This research was supported by a philanthropic gift from the Novo Nordisk corporation to the University of California, Berkeley, to support the Joint Initiative for Causal Inference.\\

\textbf{Author Contributions}

LED, MP, and MvdL developed the methods. LED wrote the code to implement the method. LED designed and ran the simulation. JBB and KK provided subject matter expertise regarding the trial used for the real data analysis. LED, JMT, and TJA designed and ran the real data analysis.  LED prepared the manuscript, with contributions from all co-authors. The authors applied the FLAE approach for the sequence of authors. All authors have accepted responsibility for the entire content of this manuscript and approved its submission.\\

\textbf{Conflict of Interest Statement}

Prof. Maya Petersen and Prof. Mark van der Laan are members of the Editorial Advisory Board in the Journal of Causal Inference, but they were not involved in the review process of this article. LED reports tuition and stipend support while at UC Berkeley from a philanthropic gift from the Novo Nordisk corporation to the University of California, Berkeley to support the Joint Initiative for Causal Inference. JMT, TJA, and KK are full-time employees of Novo Nordisk A/S and own stocks in Novo Nordisk A/S. JBB reports contracted fees and travel support for contracted activities for consulting work paid to the University of North Carolina by Novo Nordisk; grant support by Dexcom, NovaTarg, Novo Nordisk, Sanofi, Tolerion and vTv Therapeutics; personal compensation for consultation from Alkahest, Altimmune, Anji, AstraZeneca, Bayer, Biomea Fusion Inc, Boehringer-Ingelheim, CeQur, Cirius Therapeutics Inc, Corcept Therapeutics, Eli Lilly, Fortress Biotech, GentiBio, Glycadia, Glyscend, Janssen, MannKind, Mellitus Health, Moderna, Pendulum Therapeutics, Praetego, Sanofi, Stability Health, Terns Inc, Valo and Zealand Pharma; stock/options in Glyscend, Mellitus Health, Pendulum Therapeutics, PhaseBio, Praetego, and Stability Health; and board membership of the Association of Clinical and Translational Science. MvdL reports that he is a co-founder of the statistical software start-up company TLrevolution, Inc. MvdL and MP report personal compensation for consultation from Novo Nordisk.\\

\textbf{Ethical Approval}

The conducted research is not related to either human or animals use.\\

\textbf{Code and Data Availability Statement} 

The code necessary to run and analyze the simulation presented in this paper is available in the attached ``Code'' folder. This code may also be found at \url{https://github.com/Lauren-EylerDang/ESCVTMLE_JCI_Manuscript_Code}. The data sets analyzed during the current study are available from the corresponding author on reasonable request.

%\bibliographystyle{plain}
%\bibliography{reference}

\end{document}